\documentclass[a4paper,USenglish]{lipics-v2021}
\usepackage{xspace}
\usepackage{url}
\usepackage{xcolor}
\usepackage{graphicx}
\usepackage{listings}
\usepackage{hyperref}
\usepackage{subcaption}
\usepackage{booktabs}
\usepackage[font=footnotesize]{caption}
\usepackage{multirow}
\usepackage{bm,amsmath}
\usepackage{cleveref}

\nolinenumbers

\renewcommand\comment[1]{}
\newcommand\dahlia[1]{\comment{\textit{\color{olive}{Dahlia: #1}}}}
\newcommand\bano[1]{\comment{\textit{\color{blue}{Bano: #1}}}}
\newcommand\alberto[1]{\comment{\textit{\color{red}{Alberto: #1}}}}
\newcommand\todo[1]{\comment{\textit{\color{red}{TODO: #1}}}}

\ifdefined\cameraReady
\renewcommand\dahlia[1]{}
\renewcommand\bano[1]{}
\renewcommand\alberto[1]{}
\renewcommand\todo[1]{}
\fi

\newcommand\sysname{\Twins}

\newcommand\hotstuff{HotStuff\xspace}
\newcommand\fhs{Fast-HotStuff\xspace}

\newcommand\tendermint{Tendermint\xspace}
\newcommand\jepsen{Jepsen\xspace}
\newcommand\zyzzyva{Zyzzyva\xspace}
\newcommand\fab{FaB\xspace}
\newcommand\synchs{Sync HotStuff\xspace}
\newcommand\pbft{PBFT\xspace}
\newcommand\Twins{Twins\xspace}
\newcommand\TwinsImpl{LibTwins\xspace}
\newcommand\twins{twins\xspace}

\newcommand\twin{twin\xspace}
\newcommand\Byzantine{Byzantine\xspace}
\newcommand\byzantine{Byzantine\xspace}
\newcommand\bft{BFT\xspace}
\newcommand\librabft{DiemBFT\xspace}

\newcommand\playground{network playground\xspace}
\newcommand\Playground{Network playground\xspace}
\newcommand\target{compromised\xspace}
\newcommand\qc{QC\xspace}
\newcommand\qcc[1]{$QC$(#1)}

\newcommand\timeout{timeout\xspace}
\newcommand\timeouts{timeouts\xspace}
\newcommand\qcs{QCs\xspace}
\newcommand\QC{QC\xspace}
\newcommand\currentround{$current\_round$\xspace}
\newcommand\blockround{$block\_round$\xspace}
\newcommand\preferredround{$preferred\_round$\xspace}
\newcommand\parentround{$parent\_round$\xspace}
\newcommand\grandparentround{$grandparent\_round$\xspace}
\newcommand\lastvotedround{$last\_voted\_round$\xspace}
\newcommand\generator{scenario generator\xspace}
\newcommand\tester{framework\xspace}
\newcommand\executor{scenario executor\xspace}

\newcommand\testcase{scenario\xspace}
\newcommand\testcases{scenarios\xspace}

\newcommand\Testcases{Scenarios\xspace}
\newcommand\Test{Scenario\xspace}
\newcommand\test{scenario\xspace}
\newcommand\Tests{Scenarios\xspace}
\newcommand\tests{scenarios\xspace}
\newcommand\aws{AWS\xspace}
\newcommand\forcelocking{force-locking\xspace}
\newcommand\nil{Nil\xspace}
\newcommand\cc{$CC$\xspace}
\newcommand\ccc[1]{$CC$(#1)}

\newcommand{\vs}{vs.\@\xspace}

\newcommand{\etal}{et al.\@\xspace}
\newcommand{\eg}{e.g.,\@\xspace}
\newcommand{\ie}{i.e.,\@\xspace}
\newcommand{\via}{via\@\xspace}

\newcommand\smr{SMR\xspace}

\newcommand{\myrightarrow}[1]{\mathrel{\raisebox{-2pt}{$\xrightarrow{#1}$}}} 

\newcommand{\para}[1]{\vspace{2mm}\noindent\textbf{#1}.\xspace}

\def\first{({\it i})\xspace}
\def\second{({\it ii})\xspace}
\def\third{({\it iii})\xspace}
\def\fourth{({\it iv})\xspace}

\newcommand\myparagraph[1]{\vspace*{0.05in} \noindent \textbf{#1}}

\newcounter{countgap}
\definecolor{verylightgray}{gray}{0.90}
\newcounter{countinsight}
\definecolor{verylightgray}{gray}{0.90}
\lstdefinelanguage{alg}{
morekeywords={fn, let, mut, ensure, match, bail, if, None, Some, return, Result, Ok},
morecomment=[l]///,
}

\def\cameraReady{true} 

\newcommand\currentstatus{44M\xspace}

\title{\Twins: BFT Systems Made Robust}

\author{Shehar Bano}{Facebook Novi, United Kingdom}{bano@fb.com}{}{}
\author{Alberto Sonnino}{Facebook Novi, United Kingdom}{asonnino@fb.com}{}{}
\author{Andrey Chursin}{Facebook Novi, United States}{achursin@fb.com}{}{}
\author{Dmitri Perelman}{Facebook Novi, United States}{dmitrip@fb.com}{}{}
\author{Zekun Li}{Facebook Novi, United States}{zekun@fb.com}{}{}
\author{Avery Ching}{Facebook Novi, United States}{aching@fb.com}{}{}
\author{Dahlia Malkhi}{Facebook Novi, United States}{dmalkhi@fb.com}{}{}

\authorrunning{S. Bano, A. Sonnino, A. Chursin, D. Perelman, Z. Li, A. Ching and D. Malkhi}
\Copyright{Shehar Bano, Alberto Sonnino, Andrey Chursin, Dmitri Perelman, Zekun Li, Avery Ching, and Dahlia Malkhi}



\supplement{
All artifacts presented in this paper are made publicly available. Specifically, this includes: \first the Rust implementation of \TwinsImpl, the \Twins framework we implemented for DiemBFT (\Cref{implementation}); \second the artifacts (the AWS orchestration scripts, and microbenchmarking scripts and data) used to evaluate \TwinsImpl (\Cref{evaluation}); and \third the Python simulator and \Twins instantiation of safety flaw in \fhs (\Cref{known-attacks}).
}

\funding{This work is funded by Novi, a subsidiary of Facebook.}

\acknowledgements{
The authors would like to thank Ben Maurer, David Dill, Daniel Xiang, Kartik Nayak, Ling Ren, and Scott Stoller for feedback on late manuscript, and George Danezis for comments on early manuscript. We also thank the Novi Research and Engineering teams for valuable feedback.
}

\EventEditors{John Q. Open and Joan R. Access}
\EventNoEds{2}
\EventLongTitle{42nd Conference on Very Important Topics (CVIT 2016)}
\EventShortTitle{CVIT 2016}
\EventAcronym{CVIT}
\EventYear{2016}
\EventDate{December 24--27, 2016}
\EventLocation{Little Whinging, United Kingdom}
\EventLogo{}
\SeriesVolume{42}
\ArticleNo{23}

\begin{document}

\maketitle

\begin{abstract}
This paper presents Twins, an automated unit test generator of Byzantine attacks. Twins implements three types of Byzantine behaviors: (i) leader equivocation, (ii) double voting, and (iii) losing internal state such as forgetting ``locks'' guarding voted values. 
To emulate interesting attacks by a Byzantine node, it instantiates \textit{twin} copies of the node instead of one, giving both twins the same identities and network credentials. To the rest of the system, the twins appear indistinguishable from a single node behaving in a ``questionable'' manner. Twins can systematically generate Byzantine attack scenarios at scale, execute them in a controlled manner, and examine their behavior. Twins scenarios iterate over protocol rounds and vary the communication patterns among nodes. 
Twins runs in a production setting within DiemBFT where it can execute 44M Twins-generated scenarios daily. 
Whereas the system at hand did not manifest errors, subtle safety bugs that were deliberately injected for the purpose of validating the implementation of Twins itself were exposed within minutes. Twins can prevent developers from regressing correctness when updating the codebase, introducing new features, or performing routine maintenance tasks.
Twins only requires a thin wrapper over DiemBFT, we thus envision other systems using it. Building on this idea, one new attack and several known attacks against other BFT protocols were materialized as Twins scenarios. In all cases, the target protocols break within fewer than a dozen protocol rounds, hence it is realistic for the Twins approach to expose the problems.

\end{abstract}

\section{Introduction} \label{intro}
Byzantine Fault Tolerant (\bft) protocols introduced in the seminal work of Lamport~\etal~\cite{lps82} are designed to withstand attacks or arbitrary malfunction of internal nodes. However, creating \byzantine attacks in order to validate a \bft system is challenging: \first \byzantine behavior is unconstrained and \second developers may be tainted by what they think that the system is designed to tolerate. Last, as a pragmatical consideration, developing code that implements \byzantine attacks might be risky.

This paper introduces \Twins, a principled approach for effectuating \byzantine attacks on \bft systems and examining their behavior. Instead of coding incorrect behavior, \Twins creates faulty behavior testing from the correct behavior itself, simply by duplicating \textbf{correct and unmodified} node behavior. This works as follows.

Twins creates a "faulty" nodes by deploying two (or generally, $k$) instances, both having the same credentials/signing-keys but running autonomously. 
Thus, for example, both nodes can send messages in the same protocol round, but these messages will carry conflicting proposals or votes; to the rest of the system, this \twins behavior will appear indistinguishable from an equivocating behavior by a single node. 
In another example, one \twin may send a vote in one round, and its \twin will ``forget'' it has voted in the next round; again, to the rest of the system, this will appear indistinguishable from a single node violating safety rules.

\Twins is based on the insight that most interesting \byzantine attacks are internal and leverage knowledge of the expected behavior of participants, hence they go unnoticed. In particular, \Twins foregoes trivial attacks such as sending semantically invalid messages, or sending a message without justification. Thus, leveraging existing code, \Twins can automatically cover material \byzantine behaviors. Indeed, \Cref{known-attacks} demonstrates one new, and several known, attacks on BFT protocols materialized as \Twins attacks. 
Crucially, in all cases, protocols break within fewer than a dozen protocol steps, hence \Twins successfully exposes them. Note that \Twins scenarios systematically iterate over protocol rounds and vary the communication patterns among nodes. While inherently exponential, in the above attacks, it took \Twins only minutes to discover protocol flaws that in some cases, took the community decades to surface. 

\Twins has been incorporated into a production setting, DiemBFT~\cite{librabft}, in which \Twins can execute \currentstatus \Twins-generated scenarios daily. Whereas the system at hand did not manifest errors, subtle safety bugs that were deliberately injected for the purpose of validating the implementation of \Twins itself were exposed within minutes. \Twins can prevent developers from regressing correctness when updating the codebase, introducing new features, or performing routine maintenance tasks.

\myparagraph{\Twins \& attacks on BFT replication.}
\Twins arises in the context of BFT replication protocols. In this domain, several worrisome safety and
liveness vulnerabilities were exposed
recently~\cite{abraham2018revisiting,momose2020force} in both known
protocols~\cite{fab,zyzzyva} and in new ones~\cite{synchs}. 
%
One reason that BFT replication lends itself well to analysis \via \Twins is as follows.
A common paradigm underlying practical BFT replication protocols is a view-by-view design.
Each view is driven by a designated leader proposing to the nodes and going through voting rounds by the nodes.
If a leader is successful, a consensus decision is reached in the view. If not,
nodes give up after a timeout and move to the next view. Transitioning to the new
view/leader is tricky: A new leader must discover if the previous leader was
successful, but it may be able to communicate only with a subset of the nodes.
The transition logic turns out to be the source of problems in all the above
cases, hence exposing the flaw requires only one or two leader rotations.

\myparagraph{\Twins implementation.}
\Twins effectuates a \byzantine attack by a Byzantine node via instantiating \textit{twin} copies of the node instead of one, giving both twins the same identities and network credentials. To the rest of the system, the twins appear indistinguishable from a single node behaving in a ``questionable'' manner. \Twins minutely interacts with existing code to control message delivery and schedule various coarse-steps such as protocol rounds. It is practical to deploy in real systems as it uses existing node code, easily keeping up with an evolving software project.

We built an attack generator based on the \Twins approach in the \librabft
open-source project, the \bft replication core of the Diem payment system~\cite{librabft}. 
Implementing \Twins in \librabft consists of two principal parts. 

The first is a \emph{\executor} that deploys a network configuration where some nodes have
\twins. The \executor hides \twins behind a thin multiplexing wrapper; to the
rest of the system, each pair of \twins appear as a single entity. The \executor
controls the scheduling of message deliveries according to a prescribed
scenario. This is accomplished through a transport emulator in the \librabft
repository called \emph{Network Playground}. 

The second part is a \emph{\generator}. The \generator enumerates
scenarios by varying the number of nodes and the message delivery schedule, then
feeding the scenarios to the \executor. We describe in the paper several strategies 
for drastically reducing the number of scenarios through aggressive trimming of symmetrical scenarios. 
Among these strategies, one minimally ``opens'' the \librabft implementation and
lets the \executor determine when a node acts as a leader in the
consensus protocol. This removes duplicate scenarios that differ only in their
leaders. Another strategy may allow only faulty nodes to become leader.
\Cref{evaluation} reports on our experience with \Twins in \librabft.

\myparagraph{Coverage.}
What attacks does the \Twins approach capture? Developing a rigorous theory that
answers this question is an intriguing question left for future work. 
Here, we provide anecdotal evidence of coverage in three forms:

\first \Cref{scenarios} brings intuition and experience of several decades of work in the field. 
There are only a handful of ways in which a Byzantine
attacker can materially deviate from the safety rules imposed by its protocol.
For example, it can equivocate and send different proposals to different groups
of recipients, or it can pretend it did not send/receive a message and
propose or vote in a manner that conflicts with such a message. 

\second 
Evaluating within the \librabft production system \Cref{evaluation} provides compelling validation of the \Twins approach. Whereas the system at hand did not manifest errors, self-injected subtle safety bugs---for the purpose of validating the implementation of \Twins itself---were exposed within minutes. In particular, we created a simple safety-violating setting by deploying $f+1$ (instead of $f$) nodes with
\Twins, which led to an expected consistency violation within seconds. We further injected three subtle logical bugs, which only slightly deviated from the original specification. In all three cases, with only $f$ \twins (faults), \Twins successfully exposed safety violations. 

\third \Cref{new-attack} shows how \Twins can instantiate a safety violation in a new protocol described in a recent manuscript~\cite{fhs}. This highlights the importance of systematically analyzing the properties of \bft protocols using \Twins to expose subtle flaws. \Cref{known-attacks} reinstates several known attacks on BFT protocols
using the \Twins approach. These attacks cover a broad spectrum of vulnerabilities, \eg safety, liveness, timing, and responsiveness.

In some protocol steps, a node may wait for messages to determine its next action. Under \Twins, the node is forced to act according to the messages it received, as if the node provided a justification for each step in form of the history of messages it received. Deviating from this behavior was not required to reinstate any of the attacks discussed in \Cref{known-attacks}, though in principle, various deviating behaviors would not be covered by \Twins. Another coverage challenge emerges in synchronous protocols because a node behavior may be based on real time. In such protocols, \Twins essentially forces a node to behave in a timely manner. We tackle this case in one of the attacks investigated in Section~\ref{known-attacks} and demonstrate that nonetheless, a slight adaptation of the original attack reinstates the attack in \Twins. However, we do not know yet which timing attacks may not be covered. We discuss some concrete future directions in \Cref{future} for extending \Twins in the settings we explore as well as others.


\section{Motivating the \Twins Approach} \label{scenarios}
We open this section with a quick primer on the Byzantine Fault Tolerant (BFT) replication problem, and describe the notation that will be used to describe attacks through the rest of this paper. We then provide high-level intuition on why \Twins is a viable approach by showing the different kinds of \byzantine behaviors that can be captured by \Twins. (Concrete attacks using \Twins are described later in~\Cref{known-attacks} and~\Cref{validation}.)

\para{BFT Replication} \label{sec:model}
The goal of BFT replication is for a group of nodes to provide a fault-tolerant service through redundancy.
Clients submit requests to the service. 
These requests are collectively sequenced by the nodes; this enables all nodes to execute the same chain of requests and hence agree on their (deterministic) output.

Except when specifically noted, we consider protocols that maintain safety against arbitrary delays in message transmissions. That is,  we assume an \textit{asynchronous network} setting. 
The main challenge is to drive \textit{agreement} on a chain of
requests (and their output) among all non-faulty nodes despite node failures.
It is common to rely on leaders to populate the network with a unique proposal. 
During periods in which the leader is non-faulty and communication among the leader and non-faulty nodes is timely, this regime can drive consensus quickly.
This approach is called \emph{partial synchrony}, indicating that it maintains safety at all times and progress only during periods of synchrony.

In the \byzantine fault model, a node may crash or arbitrarily deviate from the protocol.
In this setting, a BFT replication system implements a fault tolerant service \via $n$ nodes, of which a threshold $f
< n/3$ may be \byzantine.
As \byzantine behavior is defined rather vaguely, there is no principled way to evaluate \bft systems.
\Twins is a new approach to systematically generate \byzantine attacks.
The main idea of \Twins is the following: running two (generally, up to $k$) autonomous instances of a node that both use correct code and share the same identity, allows us to emulate most interesting \byzantine attacks. Two nodes share the same identity when they share the same credentials and signing keys.

\para{Notation} \label{notation}
Nodes are represented by capital alphabets (\eg $A$) and the \twin of a node is represented by the same alphabet with the prime symbol (\eg $A^\prime$).
When referring to a set of nodes, we enclose them in parentheses \eg $(A,B,B^\prime)$.
We underline a node that is serving as the leader, \eg $\underline{A}$.
The adversary can delay and filter messages between nodes.
We denote partitions of nodes by enclosing them in braces, \eg $P_1=\{A, B, C, D\}$ and $P_2=\{E, F, G\}$, and reserve the capital letter $P$ to denote them.
Additionally, to show messages allowed in a given direction, we use the symbols $\to$ and $\leftrightarrow$. For example, $A \to (B,C)$ means $A$ can send messages to $B$ and $C$; similarly, $A \leftrightarrow P_2$ means $A$ can send messages to and receive messages from any node of the partition $P_2$.   
The scenarios described below use a network configuration of 7 nodes, $(A, B, C, D, E, F, G)$. \Byzantine nodes have \twins denoted with $^\prime$, as in $F^\prime$, $G^\prime$.
To experiment with any of the deviating behaviors described below, one can increase the number of \byzantine faults to $f+1$ (say $E,F,G$ have \twins $E^\prime,F^\prime,G^\prime$) and expect to see conflicting commits.


%
%
%

\para{Equivocation}
A quintessential \byzantine behavior is for a node to \emph{equivocate}. That
is, in the same step, a \byzantine node might send different messages to
different recipients.

\Twins covers equivocation by splitting honest nodes between two partitions,
each one communicating with only one \twin of each pair.
For example, 
we can split the system into
$P_1=\{A,B,C,D,\underline{F}\}, P_2=\{C,D,E,\underline{F^\prime},G\}$.
The leader(s) $F$ and $F^\prime$ execute correct leader code but nevertheless may generate conflicting proposals due to different inputs or randomness seeds. 
If there is a protocol flaw then these conflicting proposals could respectively commit in $P_1$ and $P_2$, hence safety breaks.


\para{Amnesia} 
An important role that nodes have in agreement protocols is \emph{vote} for a single proposal per view.
However, a \byzantine node might vote for a proposal and then `forget' that it has voted and vote again.  
\Twins covers amnesia by letting one of the \twins vote on one proposal. Since the other \twin is oblivious to the vote happening, it may nevertheless---albeit executing correct code---vote on a different proposal. 

More concretely, as in the scenario above, we can split the nodes into two partitions, 
$P_1=\{A,B,E,F,G\}, P_2=\{C,D,E,F^\prime,G^\prime\}$.
If there is a protocol flaw
then this double-voting behavior may result in conflicting commits in $P_1$ and $P_2$, hence safety breaks.
%
%
%

\para{Losing internal states}
Another notable deviation for \byzantine nodes is to lose their internal state, particularly a \emph{lock} that guards a value they voted for.
\Twins covers this deviation by letting one of the \twins get locked on a value in one view, but in some subsequent view, bring the other \twin who is ignorant that a lock exists.

More concretely, we can split the nodes into two partitions 
$P_1=\{A,B,E,F,G\}, P_2=\{C,D,E,F^\prime,G^\prime\}$.
In one view, the adversary relays messages only among $P_1$.
In the next view, it switches to $P_2$, causing $F^\prime,G^\prime$---albeit executing correct code---to
ignore their `previous' actions. 
This can repeat any number of times.
If there is a protocol flaw then 
conflicting proposals may commit in different views, hence safety breaks.

\section{Attacks Materialized in \Twins}

\label{known-attacks}

In this section, we demonstrate one new, and several known, attacks on \bft replication protocols, expressed as \Twins scenarios. We provide insight into the attacks and defer the details of all but the linear leader-replacement attack to an appendix, due to space constraints.


\subsection{New Attack} 

\label{new-attack}

\fhs~\cite{fhs} is a new protocol, described in a recent manuscript. It is similar to HotStuff~\cite{hotstuff-2019}, except with a 2-phase commit rule. 
The safety violation we reveal using \Twins is possible because \fhs does not require consecutive rounds in order to commit. 
Specifically, Quorum Certificates (\qcs)~\cite{hotstuff-2019} formed by some of the (partitioned) nodes do not reach the other nodes, resulting in two parallel branches that eventually commit two conflicting blocks. We instantiate this safety violation with \Twins (using only network partitions in a network with 4 nodes and within 11 rounds). This highlights the efficacy of systematically analyzing the properties of \bft protocols \via \Twins to expose subtle flaws. More details are provided in \Cref{sec:fhs}. 

We implemented the \fhs \bft consensus algorithm in a Python simulator which we release as open source\footnote{
\ifdefined\cameraReady
\url{https://github.com/asonnino/twins-simulator/tree/master/fhs}
\else
Link removed for blind review.
\fi
}. The simulator then executes \Twins scenarios over the algorithm. 


\subsection{Reinstated Attacks} 

\label{reinstated-attacks}

We present several known attacks on \bft protocols, expressed as \Twins scenarios.  
In all cases, exposing vulnerabilities requires only a small number of nodes, partitions, rounds and leader rotations.
It is worth noting that later, our evaluation (\Cref{evaluation}) of \TwinsImpl, \Twins implemented for \librabft, shows that running an automated scenario generator (\Cref{generator}) with these configurations would cover the described attacks within minutes. We did not undertake to re-implement all these protocols and apply a \Twins scenario generator to them; our implementation covers only \librabft~\cite{librabft}.


\para{Safety attack on \zyzzyva}
\zyzzyva broke new ground in BFT replication with the introduction of
an optimistic single phase ``fast track'' commit.
Eleven years elapsed from its publication until a safety flaw in \zyzzyva was discovered~\cite{abraham2018revisiting}, during which numerous research project and systems were built on it. 
\Twins generates a scenario that exposes the flaw with 4
nodes and two leader rotations: the first leader equivocates \via a \twin, and the next two leaders drop messages to/from some nodes.
The details of this attack using \Twins is described in~\Cref{zyzzyva-detail}.

\para{Liveness attack on \fab}
\fab~\cite{fab}, a precursor to \zyzzyva, is a view-based protocol with an optimistic fast track. 
Not surprisingly, a similar problem arises in \fab due to a flawed leader replacement protocol~\cite{abraham2018revisiting}, albeit manifesting as a liveness bug. \Twins exposes this bug 
in a short scenario with $n=4$ and three leader rotations, leading to a complete absence of leader proposals.
The detailed attack using \Twins is described in~\Cref{fab-detail}.

\para{Timing attack on \synchs}
\emph{Force-Locking Attack}~\cite{momose2020force} is a timing attack on a preliminary version of a synchronous BFT protocol named \synchs~\cite{synchs} (which was subsequently updated to resist the attack). As before, \Twins captures this attack with only a small system size, $n=5$, and two leader rotations. However, in order to create timing attacks, \Twins needs to be aware of timing information for protocol steps and messages deliveries. Extending \Twins
with timing data is left for future work. 
In the specific attack at hand, course-grain timing at fixed intervals---fewer than ten---suffice to reinstate the attack.
The detailed attack using \Twins is described in~\Cref{synchs-detail}.

\para{Non-Responsiveness attack on linear leader-replacement}
Practical Byzantine Fault Tolerance (\pbft)~\cite{pbft} is a seminal work that was designed to work efficiently in the asynchronous setting.
Carrying the classical PBFT solution to the blockchain world, 
\tendermint~\cite{tendermint-2018} and Capser~\cite{casper} introduced a simplified \emph{linear} strategy for leader-replacement. However, it has been observed~\cite{tendermint-2016,hotstuff-2018} that this strategy forgoes an important property of asynchronous protocols---\emph{Responsiveness}---the ability of a leader to advance as soon as it receives messages from $2f+1$ nodes.\footnote{\tendermint is a precursor to \hotstuff~\cite{hotstuff-2019} and \librabft~\cite{librabft} which operates in two-phase views, but has no Responsiveness. \hotstuff/\librabft solve this by adding a third phase.} Indeed, bringing linear leader-replacement approach into \pbft, we demonstrate a liveness attack using a \Twins scenario. Lack of progress is detected by observing that two consecutive views with honest leaders whose communication with a quorum is timely do not produce a decision.  
We present the details of this attack using \Twins in the next section.



\subsection{Non-Responsiveness Attack} \label{sec:tm}
We now describe in more detail the non-Responsiveness attack above on linear leader-replacement.
The seminal PBFT solution operates two-phase views. A simplified, linear leader-replacement works as follows. 
A leader proposes to extend the highest \emph{quorum certificate} (\qc) it knows.
A \qc is formed on a proposed value if it gathers $2f+1$ votes from nodes. 
Nodes vote on the leader proposal if it extends the highest \qc
they know.
A commit decision on the leader proposal forms if $2f+1$ nodes form a \qc, and then $2f+1$ nodes vote for the \qc. 
Progress is hinged on leaders obtaining the highest \qc from the system, otherwise liveness is broken.

Using the notation from~\Cref{notation}, the liveness attack here uses 4 replicas $(D,E,F,G)$, where $D$ has a \twin $D^\prime$.
In the first view, $D$ and $D^\prime$ generate equivocating proposals. 
Only $D, E$ receive a \qc for $D$'s proposal. 
The next leader is $F$ who proposes to 
re-propose
the proposal by $D^\prime$, which $E$ and $D$ do not vote for because they already have a \qc for that height. 
Only $F$ and $D^\prime$ receive a \qc for $F$'s proposal. 
This scenario repeats indefinitely, resulting in loss of liveness.
More specifically, this attack works as follows:

\begin{description}

\item [View 1:] Initialize $D$ and $D^\prime$ with different inputs $v_1$ and $v_2$. 
\begin{itemize}
\item Create the partitions $P_1=\{\underline{D},E,G\}$, $P_2=\{\underline{D^\prime},F\}$.

\item Let $D$ and $D^\prime$ run as leaders for one round. $D$ proposes $v_1$ to $P_1$ and gathers votes from $P_1$ creating \qcc{$v_1$}. $D^\prime$ proposes $v_2$ to $P_2$ and gathers votes but not a \qc.

\item Create the following partitions: $P_1=\{\underline{D},E\}$, $P_2=\{\underline{D^\prime},F\}$, $P_3=\{G\}$. $D$ broadcasts \qcc{$v_1$}, which only reaches $P_1$ \ie $(D,E)$.
\end{itemize}

\item [View 2:] Drop all proposals from $D$ and $D^\prime$ until View 2 starts.
\begin{itemize}
\item Remove all partitions, \ie $P=\{D,D^\prime,E,\underline{F},G\}$.

\item Let $F$ run as leader for one round. $F$ re-proposes $v_2$ (\ie $D^\prime$'s proposal in the previous round) to $P$. $(D,E)$ do not vote as they already have \qcc{$v_1$} for that height. $F$ gathers votes from the other nodes and forms \qcc{$v_2$}.

\item Create partitions $P_1=\{D,E\}$, $P_2=\{D^\prime,\underline{F}\}$, $P_3=\{G\}$.

\item $F$ broadcasts \qcc{$v_2$}, which only reaches $P_2$.
\end{itemize}

\item [View 3:] Drop all proposals from $F$ until View 3 starts.
\begin{itemize}
\item Create the partitions $P_1=\{D,\underline{E},G\}$, $P_2=\{D^\prime,F\}$.

\item Let $E$ run as leader for one round. $E$ proposes $v_3$ which extends the highest \qc it knows, \qcc{$v_1$}. As before, $E$ manages to form \qcc{$v_3$}, but as a result of a partition, the \qc will only reach $(D,E)$. Next, there is a view-change, $F$ is the new leader, and there are no partitions. $F$ proposes $v_4$ which extends \qcc{$v_2$}, the highest \qc it knows. However, $(D ,E)$ do not vote because $v_4$ does not extend their highest \qc \ie \qcc{$v_3$}. This scenario can repeat indefinitely, resulting in the loss of liveness.   
\end{itemize}
\end{description}

\section{Systematic Scenario Generation} 

\label{design}

\begin{figure}[t]
    \centering
    \includegraphics[width=0.7\textwidth]{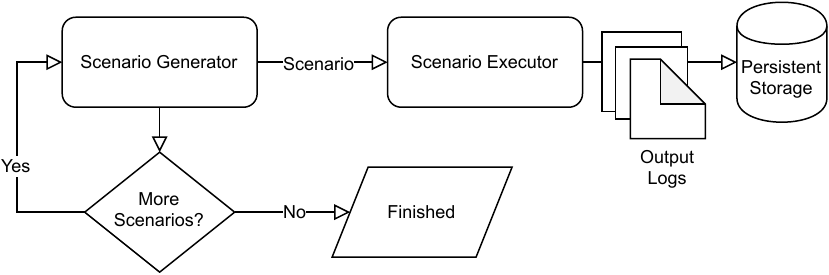}
    \caption{\Twins high-level design.}
    \label{fig:design}
\end{figure}

%
Whereas the previous section demonstrated manually crafted \Twins attack scenarios, this section presents a framework for systematically generating such scenarios.

Systematically and efficiently generating \Twins scenarios that provide good
coverage requires tailoring to the specific \bft protocol settings. 
We develop the \Twins \tester which generates and executes \emph{\testcases} that describe the node and network configurations.  
Specifically, the \Twins \tester is comprised of two components as shown in \Cref{fig:design}: \first the \executor, and \second the \generator. The \executor runs a single \testcase and generates output logs, while the \generator produces various \testcases that are fed to the \executor to check for violations. The following design goals underlie the \Twins \tester:

\begin{itemize}
    \item \textbf{Generic \& Modular.} \Twins is modular with respect to the particular \bft protocol implementation being analyzed, imposes as little complexity as possible on the development, and easily keeps up with code changes. 
    
    \item \textbf{Parametrizable.} The network setup (\ie the number of nodes, leaders per round, and network configuration per round) and adversarial assumptions (\ie how many Byzantine faults are tolerated) is configurable.
    
    \item \textbf{Feasible.} \Twins allows pruning duplicate scenarios 
     in order to provide coverage of material attacks. 
    
    \item \textbf{Customizable Coverage.} The coverage of \testcases, \ie the subset of all possible \testcases to choose for execution, is configurable by randomly sampling scenarios to run among all possible enumeration.
    
    \item \textbf{Reproducible.} \Twins writes logs to persistent storage, containing sufficient information to detect and reproduce any safety violations.
\end{itemize}

Next, we describe the two main components (\Cref{fig:design}) of \Twins---the \executor and the \generator---in detail.


\subsection{\Test Executor} \label{sec:test-executor}
\label{executor}

In every \Twins \test, a threshold of the nodes are `misconfigured' to have a \twin instance with identical transport endpoint credential and secret keys. 
%
The \Twins \executor gets as input a \testcase consisting of a node-set, a subset of which are marked \emph{\target}
(representing \byzantine nodes); 
and a round-by-round message delivery schedule. The \executor sets up a network of nodes with a given number of \target nodes 
and per round partitions and leaders. The \target nodes correspond to the nodes for which the \executor creates \twins (\ie identical instances with the same credentials and signing keys), thereby emulating misbehavior. 

As mentioned above, we address \bft replication protocols that proceed in rounds initiated by a designated leader, each round representing a state transition in the protocol's state machine replicated on each node. For each round, the \executor creates a given network partition and assigns given leaders to the round. 
The \executor runs the \bft protocol among nodes for a pre-specified number of
rounds, at the end of which, the \executor checks for violations. Specifically,
protocol guarantees can be violated in two principal ways, safety and liveness.
A safety violation is detected if two nodes commit to conflicting decisions. 
A liveness violation can be detected if the protocol fails to commit
within a certain number of steps or a certain duration bound.


\subsection{\Test Generator} \label{sec:test-generator}

\label{generator}

We build a \generator of round-by-round scenarios: for each round, the \generator enumerates possible leaders and message delivery schedules among nodes.
The \generator produces various \testcases to be fed into the \executor. Each \testcase represents a unique instance of executor configuration parameters, \ie the \target nodes and per round network partitions and leaders. 
\Testcases are generated systematically as follows (see notations in Section~\ref{notation}):  

\begin{itemize}
    
    \item \textbf{Step 1.} The \generator first produces the set of all possible partitions of nodes (called \emph{partition scenarios}). For example, for a network of 4 nodes ($A, B, C, D$), possible partition scenarios ($P$) include $\{P_1=\{A, D\}, \{B, C\} \}$, and $P_2=\{ \{A\}, \{B, C, D\} \}$. This problem relates to the \emph{Stirling Number of the Second Kind}~\cite{stirling} which enumerates the ways in which a set of $N$ nodes can be divided up into $P$ non-empty partitions, where $P$ ranges from $N$ 
    (\ie each node is self-isolated)
    to $1$ (\ie fully connected network without partitions).
    
    \item \textbf{Step 2.} Next the \generator assigns each partition scenario to all possible leaders \ie the set of $N$ nodes assuming any of those can be a potential leader. For example, for the example partition scenario above $\{ P_1=\{A, D\}, \{B, C\} \}$ for a network of nodes $(A, B, C, D)$, possible leader-partition combinations include $\{ \underline{A}, P_1 \}$, $\{ \underline{B}, P_1 \}$, $\{\underline{C}, P_1 \}$, $\{ \underline{D}, P_1 \}$. Each leader-partition combination fully describes the \Twins configuration required for each round. 
    
    \item \textbf{Step 3.} The \generator lists \testcases by enumerating all possible ways in which the leader-partition pairs generated in the previous step can be arranged over $R$ rounds (\ie permutation, with or without replacement).

\end{itemize}

\vspace{3pt}

The \generator iterates over the generated \testcases linearly, and invokes the \executor for each \testcase. For safety analysis, usually a small number of rounds ($< 10$)  
suffices to expose logical bugs in the protocol. \Test generators therefore need to enumerate a reasonable number of combinations. 

\para{Pruning \testcases} 
Important to the success of the approach is for the \generator to avoid duplicate \testcases (\eg in symmetry or node label\footnote{Nodes can have designated roles in the protocol, referred to as \emph{node labels}. \Twins incorporates the label `leader', which is the case for standard \bft protocols. Extensions of these protocols might have further hierarchy \eg primary and secondary leaders. This is currently not supported, but the \generator can be easily extended to support different node labels.} rotation) and generate only materially different scenarios. 
The implementation we describe in the Evaluation section of this paper (\Cref{evaluation}) employs aggressively such pruning. 
Certain heuristics further substantially reduce the number of scenario configurations. For example, in most safety violations the set of honest parties is split into two, hence it suffices to play with two or three partitions per round.
These optimizations make it feasible to cover a broad range of meaningful \testcases.
For analyzing liveness, many \testcases will obviously fail to make progress because there does not exist a super-majority quorum that has reliable and timely communication among its members. Hence, for liveness analysis the \generator must guarantee that eventually such a quorum exists.

\para{Message delays and \timeouts}  We note that the \generator does not address message delays and timeouts, only the dropping of messages and their relative delivery order. Because the \bft protocol may employ timers, the dropping of messages implicitly implies that relevant endpoint incur a violation of presumed bounds on transmission delays. 
Future work may incorporate explicit message delays into the \generator to check specific timing violations and also to analyze \bft protocols in the synchronous model (\Cref{future}).

\begin{figure*}[t]
    \centering
    \includegraphics[width=0.8\textwidth]{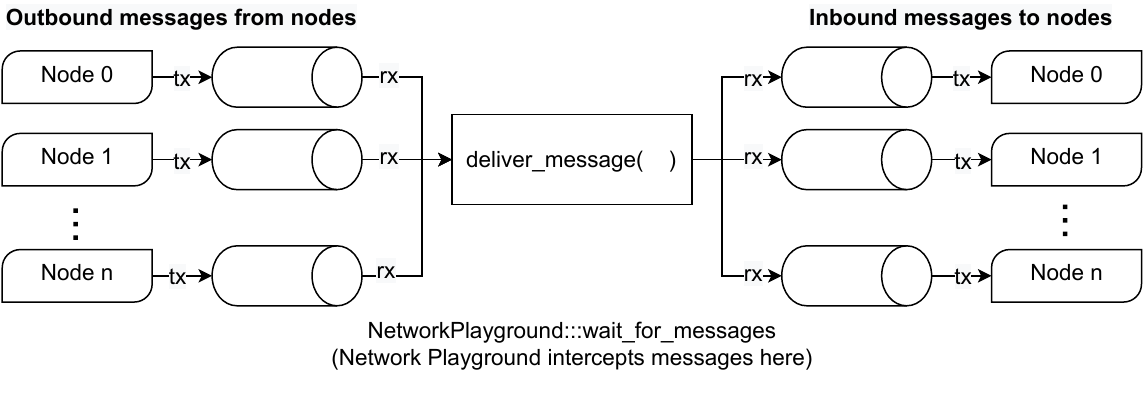}
    \caption{Design of \librabft's \emph{Network Playground}.}
    \label{fig:playground}
\end{figure*}

\section{Implementation}

\label{implementation}


We implemented the \Twins \tester for \librabft, which we call \TwinsImpl. 
\Cref{librabft} provides an overview of \librabft.
As described in \Cref{design},
an implementation consists of two principal ingredients, a \generator and an
\executor (\Cref{fig:design}). We first describe the \executor implementation which leverages a network emulator
in \librabft referred to as the \emph{\playground}. We then proceed to describe
the \generator implementation.
For completeness, the Rust code and interfaces for the main functions of \TwinsImpl,
\texttt{execute\_scenario} and \texttt{scenario\_generator}, are provided in
\Cref{sec:code}. 
We are open sourcing the Rust implementation of \TwinsImpl\footnote{
\ifdefined\cameraReady
\url{https://github.com/diem/diem}
\else
Link omitted for blind review.
\fi
}.


\subsection{\Test Executor}

\label{executor-impl}

The \TwinsImpl \executor leverages the network emulator of \librabft,
\emph{\playground}\footnote{\url{https://github.com/diem/diem/blob/master/consensus/src/network_tests.rs}}.
\Playground provides an apparatus for running single-host
\librabft deployments, emulating a network and intercepting all messages exchanged between nodes. \Tests can be written to manipulate the intercepted messages (\eg by dropping certain messages) and observe node response. Figure~\ref{fig:playground} shows the design of the \playground. Nodes are represented by processes run on different threads (that run the full consensus protocol), and network links between them are expressed as Rust channels that provide asynchronous unidirectional communication between threads. 
In \librabft, nodes are identified by their \emph{Account Address} (a public key that uniquely identifies a node). Channels are associated with their respective account addresses (nodes). When a node starts a new round, it checks whether it is leader for this round; if yes, then it generates on the fly a block to propose using a mock block generator. 
Each call to the mock block generator produces a different block. This has important implication for \TwinsImpl, as we require a node and its \twin to propose different blocks at the same round to emulate equivocation. 


The \executor component (\Cref{design}) of \TwinsImpl is built on top of \playground. This required the following modifications and extensions to the original library: 

\begin{itemize}
\item \textbf{Adding \twins.} We wrote a new method to add nodes to the network that supports \twins. The method takes `\target nodes' as a parameter to refer to the nodes for which to create \twins. For each target node, a duplicate instance is created with the same credentials and signing keys. Consequently, in the eyes of the other nodes the \target node and its twin are indistinguishable.

\item \textbf{Inferring rounds.} \TwinsImpl requires to apply a number of filtering policies at the round level. \Playground does not have a notion of rounds---it only supports static configurations that remain unchanged throughout protocol execution. There is no global notion of rounds in a distributed system with partial synchrony; instead, nodes have their own view of which round they are in, which they include in their messages. We enable \playground to extract round from intercepted messages and accordingly apply filtering criteria.  

\item \textbf{Round-based message filtering.} \Playground allows writing rules to drop intercepted messages that meet certain criteria, \ie messages to or from specified nodes and messages of specified types \eg votes or proposals. \TwinsImpl extends \playground to drop intercepted messages \emph{per round}, which allows emulating different network partitions per round. The message dropping rules treat \target nodes and their twins differently---the rules apply to account addresses (which uniquely identify nodes), not public keys (which are the same for a target node and its twins).     

\item \textbf{Deterministic multi-leader election.} \librabft currently uses a non-deterministic leader election algorithm.
\TwinsImpl requires leader election at a finer granularity, \ie assigning a specified leader to each round, potentially assigning multiple leaders to a round (because if a \target node is elected as a round leader, its \twins becomes leader too). We wrote a new leader election algorithm for \librabft that supports these requirements.   
\end{itemize}

To emulate running the protocol for a given number of rounds, we approximate rounds by the number of messages emitted by nodes.
Note that in a system with partial synchrony, we can only make guesses about rounds as there is no global notion of rounds. 
Using message-count per-round (without partitions) as an `over-guesstimate',
we let the nodes vote for $3$ extra rounds. 
Over-running a \test has no consequence on the results of \TwinsImpl (other than longer \test execution time) because any safety violations would have already been detected in earlier rounds.


\vspace{3pt}

\subsection{\Test Generator}

\label{generator-impl}

The \generator produces \testcases in three main steps. First, it generates all the possible ways in which a set of $N$ nodes can be split into $P$ partitions (partition scenarios). Second, it generates all possible ways in which $L$ leaders can be combined with the partitions generated in the previous step. Finally, it generates all the possible ways in which the partition-leader combinations can be permuted over $R$ rounds of consensus protocol execution. The \generator can operate in online or offline modes. In the online mode, \testcases are generated on the fly and fed to the \executor. The \generator can be configured to write the \testcases to a file. In the offline mode, the \generator reads previously generated \testcases from a file and feeds them to the \executor. 

\label{filters}
\para{Pruning \testcases} 
A na\"ive enumeration of all combinations of $P$ partitions, $L$ leaders, and $R$ rounds may explode quickly (see \Cref{tab:generator-stats}). 
In order to constrain the number of generated \testcases in a particular run, we provide hooks to control the number of $P$ partitions, the number of $L$ leader-partition pairs, and the number of leader-partition configuration assignments to rounds. For all three cases, we specify whether the selection is deterministic---first $X$---or randomized---an $X$ sample. In the third case---configuration assignment to rounds---the total combination space to select from is large. Therefore, the \generator allows randomizing the per-round configuration selection, rather than sampling over the entire space of assignments. 

\section{Evaluation}

\label{evaluation}

We validate the capability of \TwinsImpl to model and detect attacks, present microbenchmarks for the main components of \TwinsImpl, and describe our experiments at scale using Amazon Web Services (\aws)~\cite{aws}.
We are open sourcing the implementation of \TwinsImpl, \aws orchestration scripts, and microbenchmarking scripts and data to enable reproducible results\footnote{
\ifdefined\cameraReady
\url{https://github.com/libra/libra}
\else
Link omitted for blind review.
\fi
}.

All our evaluations correspond to 4--7 nodes, 4--7 rounds and 2--3 partitions. Intuitively, these configurations seem sufficient to expose any safety violations.     
Indeed, the known attacks on \bft protocols described in~\Cref{known-attacks} were exposed with only a small number of nodes, partitions and leader rotations. 
A recent work~\cite{Niksic2019} on the coverage of random scenarios to detect crash faults shows that coverage depends on the number of partitions and node labels (in our case, the leaders), but not on the number of nodes.
For \jepsen~\cite{jepsen}, all the bugs that provide meaningful coverage have a small number of rounds, and 2--3 partitions and roles~\cite{Niksic2019}. 
Using higher values for these parameters leads to a very large number of \testcases, which cannot be feasibly executed without some sort of filtering (Section~\ref{filters}).
It is an interesting open question whether increasing the value of these parameters has a higher chance of exposing safety violations.

\subsection{Validation} 

\label{validation}

We deliberately introduce bugs to \librabft, and validate that \TwinsImpl is able to model and detect attacks that exploit the injected vulnerabilities. 
This approach is similar to \emph{mutation testing}, a well-known technique to evaluate the quality of existing tests in terms of whether they can detect programs with deliberately injected modifications (called ``mutants''). 
While approaches such as automated mutation testing can help us to exhaustively introduce mutants, this is computationally expensive and not practical for large, complex systems. 
We select bugs to inject into \librabft based on their ability to compromise the program’s functional correctness.
We note that this choice is based on our intuition and experience, and does not provide any coverage guarantees.
The validation approach we use is to: \first inject the bug into \librabft; and \second generate \testcases using the \TwinsImpl \generator, checking for any safety violations. We instantiate the \generator with different configurations and vary them until a safety violation is exposed.

We begin with the base case: can \TwinsImpl generate a \testcase that violates safety when the \bft threshold is exceeded (\ie $>f$ \byzantine nodes)? 
We discovered a safety violation with 4 nodes and 2 \twins $(\underline{A}, B, C, D, \underline{A^\prime}, B^\prime)$, 7 rounds, and static scenario configuration (\ie each partition-leader combination is run for all $R$ rounds). \TwinsImpl executed 62 \testcases of which 8 led to safety violation within 86s.  
\bano{If we need space, the log excerpts in this section can be removed.}

\para{Changing quorum size to $\bm{2f}$}
\bft protocols consider a state transition safe if it receives votes from an honest majority of nodes (\ie quorum).
We change \librabft's quorum size from $2f+1$ to $2f$. 
\TwinsImpl detects a safety violation with 4 nodes and 1 \twin $(\underline{A}, B, C, D, \underline{A^\prime})$, 7 rounds, and static scenario configuration (\ie where each partition-leader combination is run for all the $R$ rounds). Within 20s, \TwinsImpl executes 14 \testcases of which 6 lead to safety violation.
These \testcases have the same pattern: Nodes are split into two partitions of size 2 and 3, with $A$ in one partition and $A^\prime$ in the other.    
As nodes in the two partitions can form quorum, oblivious to each other they continue to generate quorum certificates on blocks proposed by $A$ and $A^\prime$, respectively. Ultimately, nodes in the two partitions commit two different blocks. 



\para{Accepting conflicting votes}
Upon receiving a proposal, nodes vote for it only if the \blockround is greater than the \lastvotedround (Safety Rule 1, \Cref{librabft}). We introduce a subtle bug to \librabft by changing this rule, so that a node votes for a block if the \blockround is greater than \emph{or equal to} the \lastvotedround.
\TwinsImpl detects the safety violation within a few seconds, with 4 nodes and 1 twin $\{\underline{A}, B, C, D, \underline{A^\prime}\}$, and 7 rounds. 
This safety bug was detected in one-shot, with 0 partitions.
Nodes vote on proposals from both A and $A^\prime$ and quickly end up committing two different proposals for the same round.

 

\begin{figure*}[!tbp]
\centering
  \begin{subfigure}[b]{0.4\textwidth}
    \includegraphics[width=\textwidth]{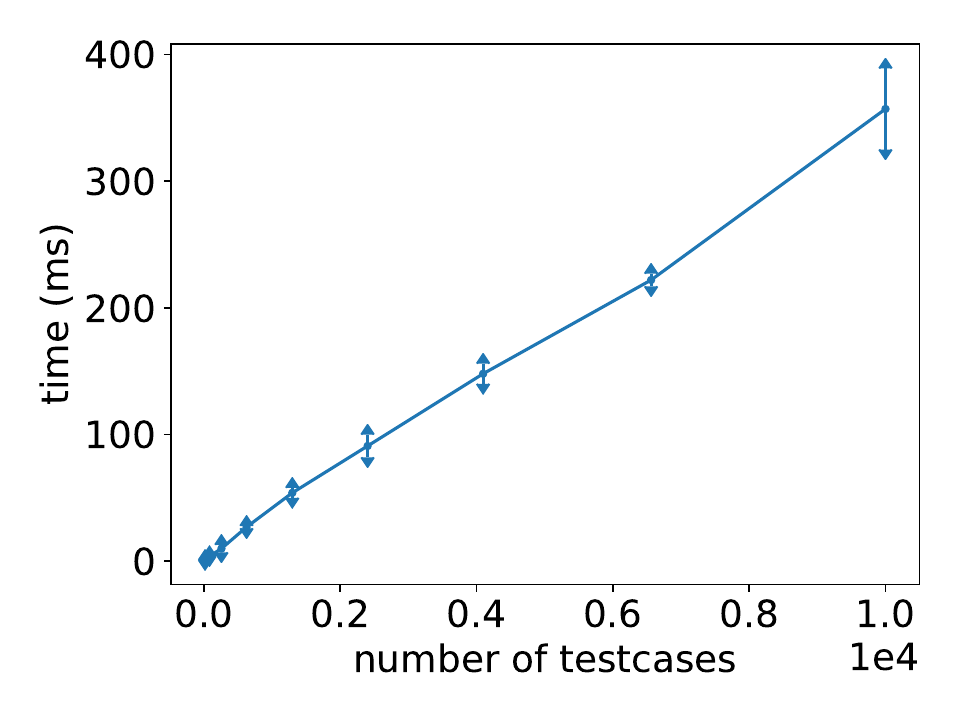}
    \caption{4 Nodes, 2 Partitions, 4 Rounds.}
    \label{fig:gentime-4}
  \end{subfigure}
  \hspace{1em}%
  \begin{subfigure}[b]{0.4\textwidth}
    \includegraphics[width=\textwidth]{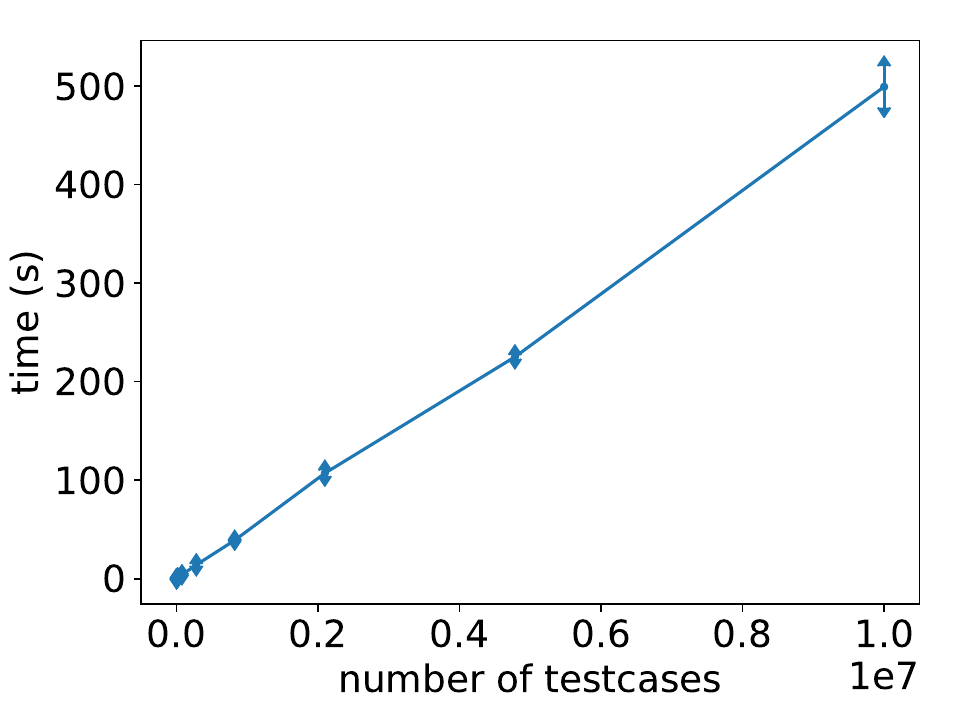}
    \caption{4 Nodes, 2 Partitions, 7 Rounds.}
    \label{fig:gentime-7}
  \end{subfigure}
  \caption{
    Time taken by the \generator to produce \TwinsImpl \testcases. Each data point is the average of 10 runs; error bars represent one standard deviation. 
    \bano{If we need space, we can remove one of the sub-figures.}
  }
  \label{fig:gentime}
\end{figure*}

\para{Forgetting to update preferred round}
Upon receiving a proposal, nodes vote for the block if the \blockround is greater than \lastvotedround, and the block's \parentround is greater than or equal to \preferredround (Safety rules 1 and 2, \Cref{librabft}). We disable the first check, and bypass the second check by never updating \preferredround so it permanently remains at 0 (Update rule 2, Section~\ref{librabft}).
The main ingredient of an attack that exploits the bug described above is to propose a block in an old round, and get the nodes to \emph{over-write} committed blocks (safety violation).
The challenge for \TwinsImpl is that as a \twin node runs correct code, it cannot be made to propose blocks in arbitrary rounds. 
One option is to partition the \twin node in an old round, and bring it back up in a later round, so it starts proposing blocks from where it left.
This is, however, not possible in a `full disclosure' protocol like \librabft where each quorum certificate (or timeout certificate) contains the full history of previous messages that led to the certificate.
That is, as soon as $A^\prime$ recovers from the partition, it receives a quorum certificate (or timeout certificate) from other nodes and advances its round. 
%
To emulate $A^\prime$ going back in time and proposing a block for an older round, we let it run as leader for a few rounds, crash it, and then recover it again as leader.
When $A^\prime$ comes back up again it starts from round 0, proposing a block that builds on the \emph{genesis} block (the first committed block). Because of our modifications to the \preferredround and \lastvotedround checks, the nodes re-write history.





\begin{table*}[t]
\centering
\footnotesize
\begin{tabular}{lrrrrrrrr} 
\toprule
\multirow{2}{*}{Nodes} & \multirow{2}{*}{Twins} & \multirow{2}{*}{Partitions} & \multirow{2}{*}{Rounds} & \multirow{2}{*}{Step 1} & \multirow{2}{*}{Step 2} &  \multicolumn{3}{c}{Step 3}\\
 &  &  &  &  &  & No Repl. & Repl. & Static \\
\midrule
4 & 1 & 2 & 4 & 15 & 15 & $\sim3\times10^{4}$ & $\sim5\times10^{4}$ & 15 \\
4 & 1 & 3 & 4 & 25 & 25 & $\sim3\times10^{5}$ & $\sim4\times10^{5}$ & 25 \\
4 & 1 & 2 & 7 & 15 & 15 & $\sim3\times10^{7}$ & $\sim2\times10^{8}$ & 15 \\
4 & 1 & 3 & 7 & 25 & 25 & $\sim2\times10^{9}$ & $\sim6\times10^{9}$ & 25 \\
\midrule
7 & 2 & 2 & 4 & 255 & 510 & $\sim7\times10^{10}$ & $\sim7\times10^{10}$ & 510 \\
7 & 2 & 3 & 4 & 3,025 & 6,050 & $\sim1\times10^{10}$ & $\sim1\times10^{15}$ & 6,050 \\
7 & 2 & 2 & 7 & 255 & 510 & $\sim9\times10^{18}$ & $\sim9\times10^{18}$ & 510 \\
7 & 2 & 3 & 7 & 3,025 & 6,050 & $\sim3\times10^{26}$ & $\sim3\times10^{26}$ & 6,050 \\
\bottomrule
\end{tabular}
\caption{The number of \TwinsImpl \testcases generated for various configurations. Steps 1, 2 and 3 correspond to the \testcase generation pipeline described in Section~\ref{design}. \textbf{Step 1:} The number of ways in which $N$ nodes can be distributed among $P$ partitions. \textbf{Step 2:} The number of ways in which the partitions generated in Step 1 can be combined with leaders. \textbf{Step 3:} The number of ways in which the partition-leader pairs generated in Step 2 can be permuted (with and without replacement) over $R$ rounds. In \textbf{Static} configurations, each partition-leader pair is statically configured for all the $R$ rounds.}

\label{tab:generator-stats}
\end{table*}

\subsection{Microbenchmarks} 

\label{microbenchmarks}

We present microbenchmarks for the two main components of \TwinsImpl: \generator (\Cref{generator-impl}) and \executor (\Cref{executor-impl}).
The microbenchmarks are run on an Apple laptop (MacBook Pro) with a 2.9 GHz Intel Core i9 (6 physical and 12 logical cores), and 32 GB 2400 MHz DDR4 RAM.

\para{\Test generator microbenchmarks}
The \generator incurs a one-time computational cost---once the \testcases are generated, the \generator feeds them one by one to the \executor. \Cref{tab:generator-stats} shows the number of \testcases generated with different configurations. We observe that the number of nodes and the number of rounds significantly increase the output of Step 1, which increases proportionally in the number of \twins (as we only configure nodes with \twins to become leaders). We find that non-static configurations in Step 3 cause the number of \testcases to explode.
Therefore, of the various filters implemented for the \generator (\Cref{filters}), we find the filter at Step 2 to be most useful. We use this filter to make our at-scale \Twins analysis (\Cref{aws}) feasible. Note that this inevitably comes at the cost of completeness of coverage---a trade-off that we cannot completely eliminate.
\Cref{fig:gentime} shows how long the \generator takes to produce \testcases for the same number of nodes (4) and partitions (2), and 4 (\Cref{fig:gentime-4}) and 7 (\Cref{fig:gentime-7}) rounds. We observe that while it expectedly takes longer to generate \testcases for 7 rounds \vs 4 rounds due to a larger number of permutations, for each case the time taken increases linearly in the number of \testcases. We observe a similar linear trend in our microbenchmarks for other configurations with varying number of nodes and partitions (figures not included due to space constraints).  

\begin{table}[!th]
\centering
\footnotesize
\begin{tabular}{lcccc} 
\toprule
\multirow{2}{*}{Rounds} & \multicolumn{2}{c}{4 Nodes} & \multicolumn{2}{c}{7 Nodes} \\
 & Mean (ms) & Std. (ms) & Mean (ms) & Std. (ms) \\
 
\midrule
4 & 239 & 314 & 547 & 1,286\\
5 & 250 & 87 & 555 & 1,059\\
6 & 284 & 88 & 555 & 802\\
7 & 296 & 87 & 559 & 752\\
8 & 334 & 209 & 647 & 810\\
9 & 363 & 175 & 643 & 557\\
10 & 398 & 222 & 653 & 539\\
11 & 433 & 168 & 718 & 570\\
12 & 465 & 179 & 748 & 223\\
\bottomrule
\end{tabular}
\caption{The time \executor takes to execute a \testcase for 4 and 7 nodes, over varying number of rounds and fixed partitions (=2). Each measurement is repeated for 100 randomly selected \testcases.}
\label{tab:executor}
\end{table}

\para{\Test executor microbenchmarks}
\Cref{tab:executor} shows the time the \executor takes to execute a \testcase. We repeat each measurement over 100 randomly selected \testcases from a configuration with 2 partitions, and varying number of nodes (4 and 7) and rounds (4--12). We observe that for 4 nodes, the execution time ranges from 234--465ms for 4--12 rounds, with a maximum standard deviation of 314ms. 
For 7 nodes, the execution time ranges from 547--748ms for 4--12 rounds, with a maximum standard deviation of $\sim 1.2$s. 

The variation observed above in execution times is expected because of how \librabft handles \timeouts (\Cref{librabft}). 
For each \testcase, \TwinsImpl runs \librabft until it has observed a given number of messages (proposals and votes), which roughly corresponds to the number of rounds. 
In some \testcases, \TwinsImpl can quickly pull out the given number of messages and finish the \testcase in a timely manner. In other \testcases, we might end up with partitions where the nodes are not able to make progress and advance rounds, due to frequent round failures and increased \timeout values. 
Some \testcases may take longer to run, waiting for the network to emit enough messages to conclude the \test. 
The execution of \testcases has negligible ($<0.1\%$) memory and CPU footprints.



\subsection{Running \Tests at Scale} 

\label{aws}

We evaluate \TwinsImpl at scale, by running it against the correct code of \librabft.  
We executed \currentstatus \testcases which were randomly selected from the 200M \testcases corresponding to the third row of \Cref{tab:generator-stats} (that is, with 4 nodes, 2 partitions, 7 rounds, permuted with replacement). 
We first generated all the 200M \testcases and randomly selected \currentstatus samples.
We ran the \generator in offline mode so the \testcases are written to file rather than being passed to the \executor.
We then split the generated \testcases into 20 shards.
The \testcases can be easily sharded, as the \tests are independent of each other---this implies that subject to the availability of computing power to generate and execute \testcases, \TwinsImpl can be scaled up arbitrarily \via sharding. 
We execute the sharded \testcases over 20 parallel instances of \TwinsImpl on \aws. We use \texttt{t3.2xlarge} instances with 8 vCPUs, 2.5 GHz, Intel Skylake P-8175; 32 GB of RAM, and 300 GB of SSD storage. All machines run a fresh installation of Ubuntu 18.04. 
We did not observe any safety violations.




\section{Related Work} 

\label{related}

There are two typical approaches to validate distributed systems. The first approach is to offer strong guarantees by building a fully verified system from the ground up~\cite{lamport:1994,prabhu2020plankton}, or to show the absence or presence of bugs~\cite{wu2014diagnosing,chen2016good,chen2014detecting,lin2013defined} by exhaustively enumerating the space of system behaviors~\cite{model-checking, yabandeh2010predicting} under systematically injected faults~\cite{LDFI, turret}.

Fully verified systems do not scale to systems deployed in the real world. Model checking and exhaustive enumeration of distributed system faults (especially, \byzantine arbitrary behavior) leads to state explosion (despite partial order reduction techniques~\cite{partial-order}), resulting in low performance. This motivates the second approach of random validation, which underlies the discipline of \emph{Chaos Engineering}, exemplified by systems like Chaos Monkey~\cite{chaos-monkey}. The main idea is to analyze the resiliency of a distributed system by randomly injecting faults (\eg terminating processes). 
Turret~\cite{turret} refines this idea by focusing on performance attacks. It runs an attack-finding algorithm using different strategies, ranging from simple brute-force to more sophisticated ``greedy search'' algorithms.
Jepsen~\cite{jepsen} is a blackbox analysis framework that runs processes with a random, auto-generated workload and randomly injected network partitions. 
A related approach is to subject the system being evaluated to \emph{trials by fire} such as Cosmos Game of Stakes~\cite{cosmos2018games}, \ie financially incentivizing the community to attack the `mock' network, and analyzing successful attacks to harden the network. Random validation is effective and scalable---but it is not comprehensive or reproducible, and cannot be used to evaluate distributed systems in an ongoing fashion.

Prior work (with the exception of \jepsen) focused on crash faults. \Twins is a new, principled approach to validate \bft systems by emulating \byzantine behavior \via \twins---copies of `compromised' nodes that can send duplicate or conflicting messages. 
\bano{If we need space, the following can be condensed to a couple of lines}
\Twins advances state-of-the-art by providing a framework to systematically generate \tests with configurable coverage, and only modeling correct executions (thus avoiding the state explosion problem associated with formal methods). We show with extensive evaluations that \Twins is suitable for evaluating real-world systems, and can be scaled up arbitrarily for larger \test coverage. \Twins automatically generate \tests that modify the interaction of components with the environment, without opening the code. 

\section{Future Work \& Conclusion}
\label{future}


\Twins is a novel approach to systematically analyze 
\bft systems. It provides coverage for many, but
not all, \byzantine attacks. The paper demonstrated anecdotal evidence of
coverage with respect to several known \byzantine attacks, and an implementation of \Twins for \librabft 
that exposes misconfiguration and purposely injected logical bugs within minutes. 
Many directions are left open for future extensions.

\myparagraph{Theory of \Twins coverage.}
As mentioned in the Introduction, it is left open to rigorously 
characterize the attacks that \Twins can cover. 
In particular, we conjecture that \Twins covers all Byzantine
behaviors in a class of protocols that have `full disclosure': each message
includes a reference to its entire causal past and any source of non-determinism (such as
local coin flips), and nodes act deterministically according to their causal
past. 
It would seem that this class of protocols is fully covered by \Twins since
the only possible attack by Byzantine nodes is to select different subsets of messages to report to
different targets.
Similarly, we conjecture that \Twins can cover timing violations in a class of
`lock-step' synchronous protocols.
Increasing coverage of \Twins in the settings we explore as well as others, and providing a formal treatment of coverage remain interesting open challenges.


\myparagraph{Checking additional properties.}
A different dimension for extension is the type of guarantees which \Twins
\tests.  While this paper focused squarely on safety of the core consensus protocol, the \Twins approach can be extended to validate ancillary components of \bft systems. For example, \librabft switches to a new set of nodes by committing a special block that includes the new set of nodes and signals the reconfiguration event. It would be useful to investigate if \Twins can cause a safety violation by creating an inconsistent node change (\ie parts of the network believe in different nodes). Similarly, \librabft's smart contract execution engine is re-instantiated \via a similar mechanism, and can be subjected to a similar \Twins-based attack.   

    
\myparagraph{Extending \Twins implementation.}
With respect to the concrete \librabft \Twins implementation presented in \Cref{implementation}, 
several extensions are left for future work, including: 
\first tackling more than a pair of \twins;
\second detecting liveness violations; and 
\third implementing process-level \twins over TCP/IP.



\bibliography{paper}

\appendix
\section{Overview of \librabft} 

\label{librabft}

We now shift our attention to utilizing \Twins for validating \bft replication in \librabft~\cite{librabft}.
We discuss our implementation and evaluation of \Twins for \librabft in~\Cref{implementation,evaluation}.
In this section, we provide an overview of \librabft (for details, see the technical report~\cite{librabft-techpaper}). 

\librabft operates in a round-by-round manner, electing leaders in each round among the nodes to balance node participation.
Rounds are slightly different from conventional ``views'' because it takes multiple rounds to
reach a decision, but leaders are rotated in each round.
The leader protocol is quite simple. A leader proposes an extension to the longest chain of requests that it knows already. Usually leaders collect batches of requests to propose, referred to as blocks, hence the \librabft protocol forms a chain of blocks (or a blockchain). 
Nodes vote for a proposed block, unless it conflicts with a longer chain that they believe may have reached consensus already. Nodes send their votes to the next leader to help the leader learn the longest safe chain.
If there are three consecutive blocks in the chain, $B_{k}$, $B_{k+1}$,
$B_{k+2}$, which are proposed in consecutive rounds, $r_{k}$, $r_{k+1}$,
$r_{k+2}$, and each block has votes from $2f+1$ nodes (gathered in a data
structure called the \emph{quorum certificate}, or \QC), then the protocol has reached consensus on block
$B_{k}$. 

If $2f+1$ send votes to the next leader in a timely manner, a \QC is formed by the
leader and it sends the next proposal. 
Nodes maintain a timer to track progress. When the timer expires
and a node still has not received a proposal, it broadcasts a timeout vote on a \nil
block. When a node gathers enough timeout votes to form a timeout certificate, it advances its
round. Every time a round fails, timeout periods are increased, allowing 
lagging nodes to catch up and enabling the protocol to eventually reach a decision.

As briefly alluded to in the Introduction, the trickiest part of BFT
replication is to manage leader transition.
\librabft maintains four parameters to ensure safety, and at the same time facilitate progress: \first \currentround, the node's current round; \second \lastvotedround, the last round for which the node voted; \third \parentround, the round of the block certified by the \QC attached with the block being processed; \fourth \grandparentround, the parent of the block certified by the \QC; and \fourth \preferredround, the highest known grandparent round. Note that as a \QC serves as a pointer to the previous certified block, \parentround and \grandparentround do not need to be explicitly tracked; these can be derived from the \QC carried by a block.       


\para{Upon Receiving a Proposal.}
Upon receiving proposal for a block, a node processes the certificates it carries, and votes for the proposed block if it satisfies a simple voting rule: If a node voted for $B_{k+2}$, it \emph{prefers} the sub-tree of proposals rooted at block $B_{k}$ (regardless of round numbers). A node will not vote for a block $B$ that does not belong to its preferred sub-tree rooted at $B_{k}$, unless $B$'s parent has votes from $2f+1$ nodes at a higher round than $r_{k}$. Concretely: 

\begin{itemize}
    \item \textbf{Safety Rule 1.} The \blockround is greater than \lastvotedround.
    \item \textbf{Safety Rule 2.} The block's \parentround is greater than or equal to \preferredround.
\end{itemize}

If the node decides to vote for the proposed block, it updates its state as follows:
    
\begin{itemize}
    \item \textbf{Update Rule 1.} Update \lastvotedround to round of the proposed block.  
    \item \textbf{Update Rule 2.} Update the node's \preferredround to the proposed block's \grandparentround if the latter is higher.
    \item \textbf{Update Rule 3.} Update the node's \currentround to the \parentround$+1$, if the latter is higher.
\end{itemize}

\para{Upon Receiving a Vote.}
For every round, the nodes send their votes to the leader of the next round. When the leader receives a vote, it performs the following safety checks:

\begin{itemize}
    \item \textbf{Safety Rule 3.} If a vote from the same node was previously received for the \emph{same} block and round, the leader rejects the vote and generates a `duplicate vote' warning.
    \item \textbf{Safety Rule 4.} If a vote from the same node was previously received for a \emph{different} block but same round, the leader rejects the vote and generates an `equivocating vote' warning.
\end{itemize}

If a vote passes both these checks, the leader considers it as valid and checks if it has enough votes to form a QC. When a QC has been formed, the leader generates a new round event, broadcasts a new block proposal and updates its state.

\begin{itemize}
    \item \textbf{Update rule 4.} When a leader gathers enough votes to form a \QC, it broadcasts a new proposal and increments \currentround.  
\end{itemize}

\emph{Spoiler alert:} In our evaluation in \Cref{validation}, we are going to deliberately modify the above rules. 
We will see that this enables safety violations that the \Twins \tester will expose.

\begin{figure}[t]
    \centering
    \includegraphics[width=.3\textwidth]{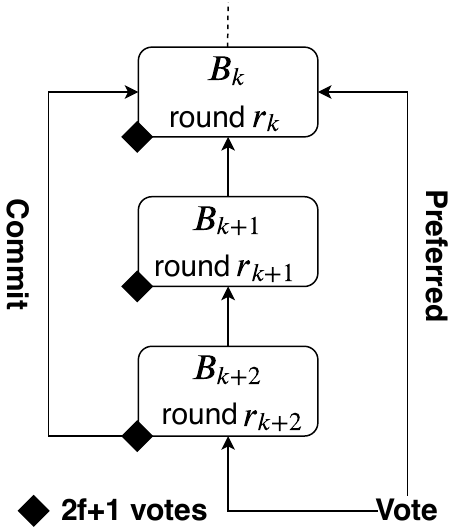}
    \caption{Consensus and preferred sub-trees in \librabft.}
    \label{fig:lbft}
\end{figure}

\section{\TwinsImpl Implementation of \Test Executor and \Test Generator}

\label{sec:code}

This section provides the Rust code for the two main functions of \Twins,
\texttt{execute\_scenario} and \texttt{scenario\_generator}.
The code listings in Figure~\ref{fig:execute-scenario} and Figure~\ref{fig:no-quorum} present simplified \Twins interfaces, \ie we omit Rust-specific features such as explicit typing, details of error messages returned,  de-referencing, and managing variable ownership.


\begin{figure}[!ht]
\begin{lstlisting}[mathescape=true]
fn execute_scenario(
    num_nodes, // number of nodes
    target_nodes, // the nodes for which to create twins
    round_partitions, // Vector of partitions for each round
    round_leaders // Vector of leaders for each round 
) {
    let runtime = consensus_runtime();
    let playground = NetworkPlayground::new(runtime.handle());

    // Start nodes and twins
    let nodes = SMRNode::start_num_nodes_with_twins(
        num_nodes,
        &target_nodes,
        &playground,
        round_proposers
    );

    // Create partitions
    create_partitions(&playground, round_partitions);

    // Start running the protocol and sending messages
    block_on(async move {
        let proposals = playground
            .wait_for_messages(2, NetworkPlayground::proposals_only::<Payload>)
            .await;

        // Pull enough votes to get a commit on the first block
        let votes: Vec<VoteMsg> = playground
            .wait_for_messages(num_nodes * num_of_rounds, NetworkPlayground::votes_only::<Payload>))
            .collect();
    });
    
    // Check that the branches are consistent at all heights
    let all_branches = vec![];
    
    for i in 0..nodes.len() {
        nodes[i].commit_cb_receiver.close();
        let node_commits = vec![];
        while let node_commit_id = nodes[i].commit_cb_receiver.try_next() {
            node_commits.push(node_commit_id);
        }
        all_branches.push(node_commits);
    }

    assert!(is_safe(all_branches));

    // Stop all nodes
    for each_node in nodes {
        each_node.stop();
    }
}
\end{lstlisting}
\vspace{-1em}
\caption{The \texttt{execute\_scenario} function which executes \testcases.}
\label{fig:execute-scenario}
\end{figure}


\begin{figure}[t]
\begin{lstlisting}[mathescape=true]

fn twins_no_quorum_scenario() {
    let runtime = consensus_runtime();
    let playground = NetworkPlayground::new(runtime.handle());
    let num_nodes = 4;
    
    // 4 honest nodes
    let n0 = 0, n1 = 1, n2 = 2, n3 = 3;
    // twin of n0
    let twin0 = node_to_twin.get(n0);
    // twin of n1
    let twin1 = node_to_twin.get(n1);
    
    // Index #s of nodes for which we will create twins
    let target_nodes = vec![0];

    // Specify round leaders
    let round_leaders = HashMap::new();
    for i in 1..10 {
        // Insert (n0, twin0, n3) as leaders for round i
        round_leaders.insert(i, vec![n0, twin0, n3]);
    }
    
    // Specify round partitions
    let round_partitions = HashMap::new();
    for r in 0..10 {
        // Insert partitions for round r
        round_partitions.insert(
            r,
            vec![
                vec![n0, twin0, n1],
                vec![n2, n3],
            ],
        );
    }
    
    execute_scenario(
        num_nodes,
        &target_nodes,
        &round_partitions,
        &round_leaders
    );
}

\end{lstlisting}
\vspace{-1em}
\caption{Twins `No Quorum' scenario.}
\label{fig:no-quorum}
\end{figure}

The \executor, implemented by $execute\_scenario$ (\Cref{fig:execute-scenario}), executes \testcases generated by the \generator. This function takes as input the number of nodes and twins, and the leaders and partitions for each round. It creates a network with the given inputs, and starts running the protocol until the nodes have emitted a given number of messages, which approximate the number of rounds for which the protocol has been run. 

The $execute\_scenario$ function exposes a simple interface, abstracting complex underlying network and SMR configurations. To demonstrate the simplicity and flexibility of $execute\_scenario$, we show how to implement a simple \test (Figure~\ref{fig:no-quorum}) where no quorum can be formed, and therefore no block gets committed. We set up a network with 4 honest nodes ($n0, n1, n2, n3$), and 1 twin ($twin0$). We split the network into two partitions $\{\underline{n0}, \underline{twin0}, n1\}$ and $\{n1, n3\}$. For each round $n0$ and $twin0$ (in partition 1) are leaders. We then run the protocol for enough rounds (at least 3 in \librabft) to get a commit on a block. In partition 1, both $n0$ and $twin0$ propose different blocks for the same rounds. $n1$ will only vote for one of the two proposals because the second proposal is for a round that is not greater than its \lastvotedround (Safety rule 1, Section~\ref{librabft}). The second partition does not have enough nodes to form quorum. Consequently, no blocks are committed.
\section{Detailed Safety Attack on \zyzzyva} 

\label{zyzzyva-detail}

We present a summary of \zyzzyva, and use \Twins to reinstate a known safety attack~\cite{abraham2018revisiting} on \zyzzyva~\cite{zyzzyva}.
We use the notation described in Section~\ref{notation}.


\subsection{Summary of \zyzzyva}

\label{zyzzyva-summary}

\zyzzyva is an SMR protocol in the same settings as \librabft (partial synchrony
and $n=3f+1$). It operates in a view-by-view manner. Each view has a designated leader. 
Nodes vote on the leader proposal if they consider it valid (we describe the validity criteria below, which has a flaw that enables the safety attack).
A commit decision on the leader proposal forms in either of two tracks,
fast and two-phase. In the fast track, all $n$ nodes vote for the leader
proposal to commit it. In the two-phase track, $2f+1$ nodes form a
commit-certificate (\cc), then $2f+1$ nodes vote for the \cc to commit
the proposal.

At the beginning of the view, nodes send the
new leader a signed NEW{-}VIEW status message. 
The leader's first proposal carries the status of $2f+1$ nodes at the beginning
of the view to prove the proposal validity.
The (flawed) definition in \zyzzyva for a valid proposal upon view change is as
follows. For each sequence slot:
\begin{itemize}
    \item \textbf{Validity Rule 1} The leader picks among the states of $2f+1$ nodes, the \cc from the highest view, if one exists.
    \item \textbf{Validity Rule 2} Otherwise, the leader picks a proposal that has $f+1$ votes from the highest view, if one exists.
    \item \textbf{Validity Rule 3} Finally, if none of the above exist, the leader creates a Nil proposal.
\end{itemize}

The flaw is to prioritize Validity Rule 1 over Validity Rule 2, which causes the leader to prefer \cc even if generated in a \emph{lower view} than $f+1$ votes.


\subsection{Safety Attack on \zyzzyva} 
\label{zyzzyva-attack}
The \zyzzyva flawed scenario safety demonstrated in~\cite{abraham2018revisiting} goes through a succession of three views. 
In the first view, a faulty leader generates conflicting proposals $v_1, v_2$ and splits honest nodes between $f+1$ that vote for $v_1$ and $f$ that vote for $v_2$. The faulty leader gathers a \cc on $v_1$ but does not send it to other nodes. 
In the second view, a good leader adopts $v_2$ and drives agreement in the fast track. 
In the third view, $f$ faulty nodes join the $f+1$ honest nodes that voted for $v_1$ in the first view. They send the leader a \cc for $v_1$,
hence the protocol proceeds with $v_1$, in conflict with the $v_2$ commit.
The attack on \zyzzyva needs only $n=4$ nodes, of which $f=1$ is faulty, and it is fairly easy to re-instate using the \Twins framework. There are four nodes, $(D, E, F, G)$. To model the case that $D$ is \Byzantine, it has a \twin $D^\prime$ initialized with different input.
We drive the execution creating partitions and electing leaders at each step, according to the attack described above.
We describe below the detailed attack using \Twins.

\begin{description}
\item [Step 1] Initialize $D$ and $D^\prime$ with different inputs $v_1$ and $v_2$. 

\item [Step 2] During View 1:

\begin{itemize}

\item Create the following partitions: $P_1=\{\underline{D},E,F\}$, $P_2=\{\underline{D^\prime},G\}$.

\item Let $D$ run as leader for one round. $D$ proposes $v_1$ to $P_1$ and gathers votes from $P_1$ creating a \cc.

\item Create the following partitions: $P_1=\{E,F\}$, $P_2=\{\underline{D^\prime},G\}$, $P_3=\{\underline{D}\}$.

\item As a result, $D$ does not get to share \cc on $v_1$ with $E$ and $F$.

\item Similarly, for one round let $D^\prime$ propose $v_2$ to $P_2$ and gather votes from $P_2$.  

\end{itemize}


\item [Step 3] Delay all messages until a new view starts. View 2:

\begin{itemize}

\item Create the following partitions: $P_1=\{D^\prime,E,\underline{G}\}$, $P_2=\{D,F\}$.

\item Run $G$ as leader, and let it collect (NEW{-}VIEW) messages from $D^\prime$ and $E$. Using Validity Rule 2 (\Cref{zyzzyva-summary}), $G$ decides to propose for $v_2$.  

\item Remove all partitions, \ie $P=\{D,D^\prime,E,F,\underline{G}\}$.

\item $G$ proposes $v_2$, and collects votes from everyone. This leads to a commit of $v_2$.

\end{itemize}


\item [Step 4] Delay all further messages until new view starts. View 3:

\begin{itemize}

\item Create the following partitions: $P_1=\{D,\underline{E},F\}$, $P_2=\{D^\prime,G\}$.

\item Run $E$ as leader, and collect (NEW{-}VIEW) messages from $D$ and $F$.  Note that $D$ sends the \cc on $v_1$ (from view 1) to $E$. Using Validity Rule 1 (\Cref{zyzzyva-summary}), $E$ decides to propose $v_1$. 

\item $E$ proposes $v_1$ to $P_1$, and gathers votes from $D$, $E$ and $F$ (who empty their local logs, undoing $v_2$). This leads both $E$ and $F$ to commit $v_1$, a safety violation. 

\end{itemize}

\end{description}

\section{Detailed Liveness Attack on \fab} 

\label{fab-detail}

We present a summary of \fab, and use \Twins to reinstate a known liveness attack on \fab~\cite{abraham2018revisiting}.
We use the notation described in Section~\ref{notation}. 


\subsection{Summary of \fab} 
\label{fab-summary}
\fab is a single-shot consensus protocol for the partial synchrony setting with
$n=3f+1$.\footnote{FaB is actually designed for a \emph{parameterized} model with
$n=3f+2t+1$, with safety guaranteed against $f$ Byzantine failures and fast track
guaranteed against $t$. For brevity and uniformity, we ignore $t$ here and set
$t=0$.}

A precursor to \zyzzyva, \fab is a view-based protocol with an optimistic fast
track. A leader drives a decision in the fast track if all nodes vote for it,
and in the two-phase track if $2f+1$ nodes vote for a $(2f+1)$
commit-certificate (\cc). When a new leader is elected, it picks a valid proposal that does not conflict with neither $f+1$ votes nor a \cc in the previous view.


\subsection{Liveness Attack on \fab} 
\label{sec:fab-attack}
The (flawed) selection criterion above leads an execution in the following scenario to become stuck. A faulty leader equivocates and proposes $v_1, v_2$ to $2f+1$ and
$f$ honest nodes, respectively. In transitioning to the next view, there is
a commit-certificate for $v_1$ and $f+1$ votes 
for $v_1$ (including an equivocation by one faulty), hence neither is safe,
and the new leader is stuck.
The attack on \fab needs only $n=4$ nodes, of which $f=1$ is faulty, and it can be easily re-instated using \Twins. There are four nodes, $(A, B, C, D)$ with $D$ as a \Byzantine node, for which we create a \twin $D^\prime$ initialized with different input.
We describe below the attack using \Twins.


%



\begin{description}


\item [Step 1] Initialize $D$ and $D^\prime$ with different inputs $v_1$ and $v_2$. 


\item [Step 2] During View 1:

\begin{itemize}

\item Create the following partitions: $P_1=\{A,B,\underline{D}\}$, $P2=\{C,\underline{D^\prime}\}$

\item Run $D$ as leader for one round. $D$ proposes $v_1$ to $P_1$ which decides to vote on $v_1$.

\item Insert the following rule in $P_1$: $(B,D) \to A$. That is, the only messages allowed are those from $B$ and $D$, to $A$. 

\item $D$, $A$ and $B$ send their votes which only reach $A$. Thus, only $A$ produces a \cc for $v_1$.

\item Meanwhile, the leader $D^\prime$ proposes $v_2$ to $P_2$.
\end{itemize}


\item [Step 3] Delay all further messages until new view starts. Create the partitions: $\{\underline{A},C,D^\prime\}$, $\{B,D\}$. Let the new leader $A$ collect NEW{-}VIEW status messages from $P_1$. These status messages block $A$  from proposing both $v_1$ and $v_2$ due to the \fab proposal validity rule. The rule states that a proposal is valid if it does not conflict with neither $f+1$ votes nor a \cc in the previous view, which is not the case for $v_1$ (has a \cc) and $v_2$ (has $f+1$ votes) as described below:

\begin{itemize}

\item From $A$, the NEW{-}VIEW message contains the value $v_1$, and a \cc for it.

\item From $C$, the NEW{-}VIEW message contains the value $v_2$, and no \cc.

\item From $D^\prime$, the NEW{-}VIEW message contains the value $v_2$, and no \cc.

\end{itemize}

\end{description}

\section{Detailed Liveness Attack on \synchs} 

\label{synchs-detail}

We present a summary of \synchs, and use \Twins to reinstate the \forcelocking attack~\cite{momose2020force} on a preliminary version of \synchs (which was fixed in an updated version). 
We use the notation described in Section~\ref{notation}.

\subsection{Summary of \synchs} 
\label{synchs-summary}
The preliminary version of~\synchs~\cite{synchs} is an \smr solution in the synchronous model with $n=2f+1$ parties.\footnote{The description here covers the first of three 
variants in that paper; two other variants 
are designed for slightly different synchrony assumptions, but the attacks
on them are similarly covered by the \Twins approach.} 

In synchronous protocols like \synchs, nodes execute the protocol in terms of $\Delta$, which is the known bound assumed on maximal network transmission delay.
\synchs operates in a view-by-view regime 
---in each view there is a designated leader which proposes values to nodes.
If a node accepts the proposed value, it broadcasts its vote.
A node creates a commit certificate (\cc) for a proposed value if it receives $f+1$ votes on it.
Nodes track the highest \cc, and only vote on a proposed value if it: \first extends the highest \cc known to the node, and \second does not equivocate another value proposed for the same height.

A node creates and broadcasts a \emph{blame} against a leader: \first if the leader does not propose a value for $3\Delta$, or \second the leader proposes an equivocating value. 
If a node observes $f+1$ blames against the leader in the current view, it broadcasts the $f+1$ blames, then waits $\Delta$ (to allow the blames to reach all honest nodes), and moves to the new view. 
In the new view, it immediately sends the new leader
the highest \cc it knows of.

After a view change, the new leader waits 
for $\Delta$ to receive node status messages (carrying the highest \cc known to them).
The leader then proposes a value that extends the highest \cc from among the received status messages.
Nodes proceed in the new view as previously described.


\subsection{Implementing Synchrony Attacks in \Twins} 
Due to the synchronous settings and the nature of the attack which heavily
leverages synchrony assumptions, in this case a \Twins scheduler must control message delivery timing. 
More precisely, rather than only specifying whether a message is delivered to a party or dropped, attacks on synchronous protocols require the \Twins scheduler to deliver messages to specific parties at specified times.
While this is captured by the \Twins approach, our current implementation  (Section~\ref{implementation}) does not support this feature (this will be implemented in future \Twins extensions).

Generally, we expect that the granularity of the scheduler timing can be fairly coarse. In particular, there is a known
parameter $\Delta$, the bound presumed by the algorithm on message transmission delays and hard-coded into it. 
Indeed, the \forcelocking attack needs to deliver messages at $0.5 \Delta$ increments, \eg at times $0, 0.5\Delta, \Delta, 1.5\Delta,
2.0\Delta, ..$. Therefore, a \Twins network emulator could operate in discrete lock-step at $0.5\Delta$ increments.
With this capability in place, the \forcelocking attack can be re-instated in the \Twins approach as described below. 


\subsection{Safety Attack on \synchs} 
\label{synchs-attack}
We now rebuild the \forcelocking attack on the preliminary version of \synchs using \Twins.
The crux of the attack is for a faulty leader to generate a last-minute
proposal that reaches only half of the honest nodes. 
The other half trigger a view change, and now the system becomes split. 
The first half continues to commit the first leader proposal with ``help'' from \byzantine nodes. The second half starts a new view and fork the chain.
This attack can be reinstated with \Twins using 5 nodes $(A, B, C, D, E)$, of which $(A, B)$ are faulty and have \twins $(A^\prime, B^\prime)$. 

\para{Notation} We extend the notation described in Section~\ref{notation} to capture message transmission in the synchronous setting as follows: $S_t \myrightarrow{v} {S^\prime}_{t^\prime}$ denotes the transmission of a value $v$ from a set of nodes $S$ that generate the value at time $t$, to a set of nodes $S^\prime$ that receive the value at time $t^\prime$.
If a value is broadcast, we use the $\star$ symbol instead of a set: For example, $S_t \myrightarrow{V} \star$ means that $S$ broadcasts a value $v$ at time $t$.
Additionally, to highlight the `send' or `receive' action on a value, we use bold text on the left or right side of the arrow, respectively.
For example, ${\bm{S}_{\bm{t}}} \myrightarrow{v} S^\prime$ means that $S$ sends $v$ to $S^\prime$ (message arrival time is not known). 

%









To reinstate this attack with \Twins, we deploy 5 nodes $(A, B, C, D, E)$, of which $(A, B)$ are faulty and have \twins $(A^\prime, B^\prime)$. Here, $n=5$, $f=2$, and quorum size is $3$ (since synchronous \bft protocols tolerate $f$ \byzantine nodes for $n=2f+1$). We describe below the detailed attack using \Twins.

\begin{description}
\item [At time $\bm{1.5\Delta}$]: 
\begin{itemize}

\item $A$ is the leader, and broadcasts a proposal with $delay=\Delta$ for the value $v_1$ which extends $v_0$.
	
	\begin{itemize}
	\item[] $\bm{(A)}_{\bm{1.5\Delta}} \myrightarrow{propose(v_1)} \star$
	\end{itemize}

\end{itemize}

\item [At time $\bm{2.5\Delta}$]: 
\begin{itemize}

\item $C$ receives $V_1$, and broadcasts its vote.

	\begin{itemize}
	\item[] $(A)_{1.5\Delta} \myrightarrow{propose(v_1)} \bm{(C)}_{\bm{2.5\Delta}}$
	\item[] $\bm{(C)}_{\bm{2.5\Delta}} \myrightarrow{vote(v_1)} \star$
	\end{itemize}

\end{itemize}

\item [At time $\bm{3\Delta}$]:
\begin{itemize}

\item $D$ blames $A$ since it did not receive a proposal from $A$ within $3\Delta$. \Twins $(A^\prime,B^\prime)$ also did not receive a proposal from $A$, hence they also blame with $A$. $(D,A^\prime,B^\prime)$ broadcast their blames with $delay=0$, receive $f + 1$ blames from each other, and start waiting for $\Delta$.
	
	\begin{itemize}
	\item[] $\bm{(D,A^\prime,B^\prime)}_{\bm{3\Delta}} \myrightarrow{blame(A)} \star$
		\item[] $(D,A^\prime,B^\prime)_{3\Delta} \myrightarrow{blame(A)} \bm{(D,A^\prime,B^\prime)}_{\bm{3\Delta}}$
	\end{itemize}

\end{itemize}

\item [At time $\bm{3.5\Delta}$]: 
\begin{itemize}

\item $D$ receives $C$'s vote on $v_1$, but it cannot create a \cc on $v_1$ since it has less than $f + 1$ votes. 
	
	\begin{itemize}
	
	\item[]  $(C)_{2.5\Delta} \myrightarrow{vote(v_1)} \bm{(D)}_{\bm{3.5\Delta}}$
	
	\end{itemize}
	
\item $(A,B)$ broadcast their votes on $v_1$, which arrive at $C$ with delay 0. As a result, $C$ gathers $f + 1$ votes on $v_1$ and creates \ccc{$v_1$}.

	\begin{itemize}
	
	\item[] $\bm{(A,B)}_{\bm{3.5\Delta}} \myrightarrow{vote(v_1)} \star$
	
	\item[] $(A,B)_{3.5\Delta} \myrightarrow{vote(v_1)} \bm{(C)}_{\bm{3.5\Delta}}$

	\end{itemize}

\end{itemize}

\item [At time $\bm{4\Delta}$]: 
\begin{itemize}
\item $C$ receives $f+1$ blame messages from $(D,A^\prime,B^\prime)$, broadcasts all blame messages, and starts waiting for $\Delta$.

    \begin{itemize}

        \item[] $(D,A^\prime,B^\prime)_{3\Delta} \myrightarrow{blame(A)} \bm{(C)}_{\bm{4\Delta}}$
        
        \item[] $\bm{(C)}_{\bm{4\Delta}} \myrightarrow{blame(A)} \star $

	\end{itemize}
\item $D$ has waited for $\Delta$ since it quit the old view $w$ with leader $A$, so it starts the next view $w + 1$ and sends its highest commit certificate \ccc{$V_0$} along with $f+1$ blames on $A$ to the next leader $B$, with $delay=0$. 

    \begin{itemize}

        \item[] $\bm{(D)}_{\bm{4\Delta}} \myrightarrow{CC(v_0), blame(A)} (B)_{4\Delta}$

	\end{itemize}

\item  The new leader $B$ receives \ccc{$v_0$} from $D$ and $f+1$ blames on $A$, and broadcasts a proposal for value ${v_1}^\prime$ extending $V_0$. Note that $B$ does not know about \ccc{$v_1$}.

    \begin{itemize}
    
        \item[] $(D)_{4\Delta} \myrightarrow{CC(v_0), blame(A)} \bm{(B)}_{\bm{4\Delta}}$
        
        \item[] $\bm{(B)}_{\bm{4\Delta}} \myrightarrow{propose({v_1}^\prime)} \star $

	\end{itemize}
\item  $D$ receives the proposal ${v_1}^\prime$ from $B$, and broadcasts its vote with delay $\Delta$, then it sets its commit timer to $2\Delta$ and starts counting down.

    \begin{itemize}
        
        \item[] $(B)_{4\Delta} \myrightarrow{propose({v_1}^\prime)} \bm{(D)}_{\bm{4\Delta}}$
        
                \item[] $\bm{(D)}_{\bm{4\Delta}} \myrightarrow{vote({v_1}^\prime)} \star$

	\end{itemize}
\end{itemize}

\item [At time $\bm{4.5\Delta}$]: 
\begin{itemize}
\item $D$ receives votes on $v_1$ from $(A,B)$; as it has now gathered $f + 1$ votes on $v_1$ it creates \ccc{$v_1$}. However, this certificate is too late, as we will see in the following steps.

    \begin{itemize}
    
        \item[] 
        	\item[] $(A,B)_{3.5\Delta} \myrightarrow{vote(v_1)} \bm{(D)}_{\bm{4.5\Delta}}$
        	
	\end{itemize}
\end{itemize}

\item [At time $\bm{5\Delta}$]: 
\begin{itemize}
\item $C$ has waited for $\Delta$ since it quit the old view with leader $A$, so it starts the next view $w + 1$ and sends its highest certificate \ccc{$v_1$} to the new leader $B$.

    \begin{itemize}
    
        \item[] $\bm{(C)}_{\bm{5\Delta}} \myrightarrow{CC(v_1)} (B)$
        	
	\end{itemize}
	
\item $C$ receives $D$'s vote on ${v_1}^\prime$ but does not vote since ${v_1}^\prime$ (which extends \ccc{$v_0$}) does not extend its highest certificate \ccc{$V_1$}.

    \begin{itemize}
    
        \item[] $(D)_{4\Delta} \myrightarrow{vote({v_1}^\prime)} \bm{(C)}_{\bm{5\Delta}}$
        	
	\end{itemize}
\end{itemize}

\item [At time $\bm{6\Delta}$]: 
\begin{itemize}
\item $D$ commits ${v_1}^\prime$ since it finished waiting for $2\Delta$ and observed no equivocation or blame in the view $w+1$. However, $D$'s highest certificate is \ccc{$v_1$} (see time $4.5\Delta$). 

\item Now if the current leader $B$ goes offline, this will result in a view change to view $w+2$ and the new leader will extend the blockchain from the highest certificate from the previous view, \ccc{$v_1$}. But $D$ has committed ${v_1}^\prime$ conflicting with $v_1$, hence safety is violated.

\end{itemize}

\end{description}

\section{Attack on \fhs} \label{sec:fhs}
We present a safety attack against Fast-HotStuff~\cite{fhs} and express it using \sysname.

\subsection{Summary of \fhs}
\fhs is essentially HotStuff~\cite{hotstuff-2019} with a 2-phase commit rule. In the happy-path, if the leader of round $n$ is successful, the leader of round $n+1$ performs the same protocol as HotStuff, namely, it collects a QC from previous round and embeds it in the $n+1$ proposal.
In the unhappy-path, if the leader of round $n$ is unsuccessful, the protocol for leader  $n+x+1$ ($x > 0$) provides a proof in the $n+x+1$ proposal that it is using the highest QC from $2f+1$ validators. This proof incurs quadratic communication complexity. Moreover, \fhs claims it does not require consecutive rounds in order to commit.

The benefits of \fhs are twofold. It provides a fast 2-phase track for HotStuff whenever the leader is successful in obtaining a QC for the previous round (happy-path). \fhs is also faster both in phases (2 phases instead of 3) and in getting to a scenario that guarantees progress, namely, it requires 3 consecutive honest leaders (instead of 4). Requiring a leader proof for the unhappy-path prevents a proposal that conflicts with an uncommitted and unlocked tail of a chain that already has a QC. Thus, dishonest leaders cannot intentionally slow down progress by overriding the latest tail.   

\fhs is however flawed as explained in \Cref{sec:fhs-attack}.

\subsection{Safety Attack on \fhs} \label{sec:fhs-attack}
\begin{figure}[t]
\includegraphics[width=0.45\textwidth]{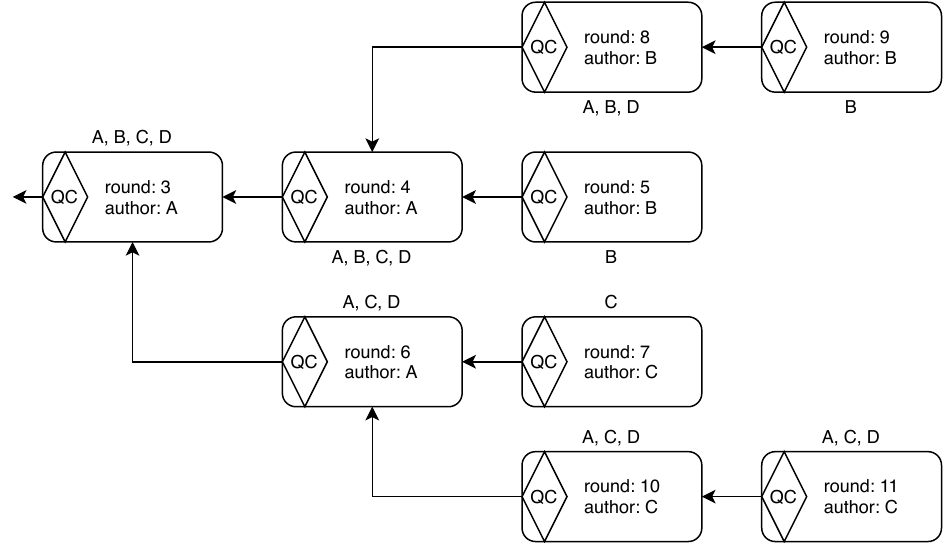}
\caption{Example of safety attack on \fhs.}
\label{fig:fhs}
\end{figure}

\Cref{fig:fhs} illustrates the safety attack against \fhs that we implement using \sysname.
There are four nodes $(A, B, C, D)$ all of which are honest---the safety attack can be executed leveraging only network partitions. Blocks are represented by rectangles (which are annotated with the nodes that receive the block). Block proposers are indicated as `authors'. Diamonds refers to QCs (which are embedded into blocks). The arrows indicate the block that a QC refers to.

We execute the safety attack in 11 rounds starting at round 3 (rounds 2 carries QC for the genesis block). 
\begin{description}
\item [Round 3:] Initially there are no partitions, \ie $P=\{\underline{A},B,C,D\}$.
\begin{itemize}
    \item $A$ proposes a block. Nodes send their votes on this proposal to the leader of the next round, node $A$.
\end{itemize}

\item[Round 4:] \phantom{*Formatting trick*--this text will be invisible}
\begin{itemize}
    \item $A$ gathers votes from the previous round, forms a QC, and includes the QC in a new block proposal. Nodes send their votes on the new proposal to the leader of the next round, node $B$.
\end{itemize}

\item[Round 5:] Set node $B$ as leader, \ie $P=\{A,\underline{B},C,D\}$.  
\begin{itemize}
    \item $B$ gathers votes from the previous round, forms a QC, and includes the QC in a new block proposal.
    \item Create the following partitions: $P_1=\{A, C, D\}$ and $P_2=\{\underline{B}\}$.
    \item The partitions prevent $B$ from broadcasting the new block (and the newly formed QC it embeds). $B$ is thus the only node knowing the QC certifying the block of round 4.
    \item Nodes of $P_1$ time out, and send a NEW{-}VIEW message to the leader of the next round (node $A$) containing their highest known QC.
\end{itemize}

\item [Round 6:] Set node $A$ as leader, \ie $P_1=\{\underline{A}, C, D\}$ and $P_2=\{B\}$.
\begin{itemize}
    \item $A$ selects the highest QC from the NEW{-}VIEW messages (\ie the QC certifying the block of round 3), and embeds it in a new block proposal. All nodes of $P_1$ vote on this proposal and send their votes to the leader of the next round (node $C$).
\end{itemize}

\item [Round 7:] Set node $C$ as leader, \ie $P_1=\{A, \underline{C}, D\}$ and $P_2=\{B\}$.
\begin{itemize}
    \item $C$ gathers votes from the previous round, forms a QC, and includes the QC in a new block proposal.
    \item Create partitions: $P_1=\{A, B, D\}$ and $P_2=\{\underline{C}\}$.
    \item These partitions prevent $C$ from broadcasting the new block (and the newly formed QC it embeds). $C$ is thus the only node knowing the QC certifying the block of round 6.
    \item Nodes of $P_1$ time out and send a NEW{-}VIEW message to the leader of the next round (node $B$) containing their highest known QC.
\end{itemize}

\item [Round 8:] Set node $B$ as leader, \ie $P_1=\{A, \underline{B}, D\}$ and $P_2=\{C\}$.
\begin{itemize}
    \item $B$ selects the highest QC from the NEW{-}VIEW messages (\ie the QC certifying the block of round 4, presented by $B$), and embeds it in a new block proposal. All nodes vote on this proposal and send their votes to the leader of the next round (node $B$).
\end{itemize}

\item [Round 9:] \phantom{*Formatting trick*--this text will be invisible}
\begin{itemize}
    \item $B$ gathers all votes from the previous round, forms a QC, and includes the QC in a new block proposal.
    \item Create partitions $P_1=\{A, C, D\}$ and $P_2=\{\underline{B}\}$.
    \item The partitions prevent $B$ from broadcasting its newly block(and the newly formed QC it embeds). $B$ is thus the only node knowing the QC certifying the block of round 8 and committing the block at round 4.
    \item Nodes of $P_1$ time out and send a NEW{-}VIEW message to the leader of the next round (node $C$) containing their highest known QC.
\end{itemize}

\item [Round 10:] Set node $C$ as leader, \ie $P_1=\{A, \underline{C}, D\}$ and $P_2=\{B\}$.
\begin{itemize}
    \item $C$ selects the highest QC from the NEW{-}VIEW messages from the previous round (the QC certifying the block of round 6, presented by $C$), and embeds it in its new block proposal. The highest QC in the NEW{-}VIEW messages.
    \item All nodes of $P_1$ vote on this proposal and send their votes to the leader of the next round (node $C$).
\end{itemize}

\item [Round 11:] Set node $C$ as leader, \ie $P_1=\{A, \underline{C}, D\}$ and $P_2=\{B\}$.
\begin{itemize}
    \item $C$ assembles votes from the previous round into a QC certifying the block of round 10, thus committing the block of round 6.
\end{itemize}
\end{description}
The safety violation appears at round 11 when node $C$ commits the block of round 6 while node $B$ previously committed the block of round 4: both blocks have the same height and fork from the block of round 3.

\subsection{Implementation of the Attack}
We implemented a Python simulator of \fhs using the discrete event simulator \emph{simpy}. We demonstrate the safety violation by running a manually-crafted scenario in the simulator. We are open sourcing our \fhs simulator as well as our \sysname scenario used for the attack\footnote{
\ifdefined\cameraReady
\url{https://github.com/asonnino/twins-simulator}
\else
Link removed for blind review.
\fi
}.

\end{document}